%
%
%

\ifx\mnmacrosloaded\undefined 
%
%
%
%

\catcode `\@=11 

\def\@version{1.6}
\def\@verdate{18th September 1995}

%
%


\newif\ifprod@font

\ifx\@typeface\undefined
  \def\@typeface{Comp. Modern}\prod@fontfalse
\else
  \prod@fonttrue 
\fi

\def\newfam{\alloc@8\fam\chardef\sixt@@n} 

\ifprod@font
\font\fiverm=mtr10 at 5pt
\font\fivebf=mtbx10 at 5pt
\font\fiveit=mtti10 at 5pt
\font\fivesl=mtsl10 at 5pt
\font\fivett=cmtt8 at 5pt     \hyphenchar\fivett=-1
\font\fivecsc=mtcsc10 at 5pt
\font\fivesf=mtss10 at 5pt
\font\fivei=mtmi10 at 5pt      \skewchar\fivei='177
\font\fivesy=mtsy10 at 5pt     \skewchar\fivesy='60

\font\sixrm=mtr10 at 6pt
\font\sixbf=mtbx10 at 6pt
\font\sixit=mtti10 at 6pt
\font\sixsl=mtsl10 at 6pt
\font\sixtt=cmtt8 at 6pt      \hyphenchar\sixtt=-1
\font\sixcsc=mtcsc10 at 6pt
\font\sixsf=mtss10 at 6pt
\font\sixi=mtmi10 at 6pt       \skewchar\sixi='177
\font\sixsy=mtsy10 at 6pt      \skewchar\sixsy='60

\font\sevenrm=mtr10 at 7pt
\font\sevenbf=mtbx10 at 7pt
\font\sevenit=mtti10 at 7pt
\font\sevensl=mtsl10 at 7pt
\font\seventt=cmtt8 at 7pt     \hyphenchar\seventt=-1
\font\sevencsc=mtcsc10 at 7pt
\font\sevensf=mtss10 at 7pt
\font\seveni=mtmi10 at 7pt      \skewchar\seveni='177
\font\sevensy=mtsy10 at 7pt     \skewchar\sevensy='60

\font\eightrm=mtr10 at 8pt
\font\eightbf=mtbx10 at 8pt
\font\eightit=mtti10 at 8pt
\font\eighti=mtmi10 at 8pt      \skewchar\eighti='177
\font\eightsy=mtsy10 at 8pt     \skewchar\eightsy='60
\font\eightsl=mtsl10 at 8pt
\font\eighttt=cmtt8             \hyphenchar\eighttt=-1
\font\eightcsc=mtcsc10 at 8pt
\font\eightsf=mtss10 at 8pt

\font\ninerm=mtr10 at 9pt
\font\ninebf=mtbx10 at 9pt
\font\nineit=mtti10 at 9pt
\font\ninei=mtmi10 at 9pt      \skewchar\ninei='177
\font\ninesy=mtsy10 at 9pt     \skewchar\ninesy='60
\font\ninesl=mtsl10 at 9pt
\font\ninett=cmtt9             \hyphenchar\ninett=-1
\font\ninecsc=mtcsc10 at 9pt
\font\ninesf=mtss10 at 9pt

\font\tenrm=mtr10
\font\tenbf=mtbx10
\font\tenit=mtti10
\font\teni=mtmi10		\skewchar\teni='177
\font\tensy=mtsy10		\skewchar\tensy='60
\font\tenex=cmex10
\font\tensl=mtsl10
\font\tentt=cmtt10		\hyphenchar\tentt=-1
\font\tencsc=mtcsc10
\font\tensf=mtss10

\font\elevenrm=mtr10 at 11pt
\font\elevenbf=mtbx10 at 11pt
\font\elevenit=mtti10 at 11pt
\font\eleveni=mtmi10 at 11pt      \skewchar\eleveni='177
\font\elevensy=mtsy10 at 11pt     \skewchar\elevensy='60
\font\elevensl=mtsl10 at 11pt
\font\eleventt=cmtt10 at 11pt     \hyphenchar\eleventt=-1
\font\elevencsc=mtcsc10 at 11pt
\font\elevensf=mtss10 at 11pt

\font\twelverm=mtr10 at 12pt
\font\twelvebf=mtbx10 at 12pt
\font\twelveit=mtti10 at 12pt
\font\twelvesl=mtsl10 at 12pt
\font\twelvett=cmtt12             \hyphenchar\twelvett=-1
\font\twelvecsc=mtcsc10 at 12pt
\font\twelvesf=mtss10 at 12pt
\font\twelvei=mtmi10 at 12pt      \skewchar\twelvei='177
\font\twelvesy=mtsy10 at 12pt     \skewchar\twelvesy='60

\font\fourteenrm=mtr10 at 14pt
\font\fourteenbf=mtbx10 at 14pt
\font\fourteenit=mtti10 at 14pt
\font\fourteeni=mtmi10 at 14pt      \skewchar\fourteeni='177
\font\fourteensy=mtsy10 at 14pt     \skewchar\fourteensy='60
\font\fourteensl=mtsl10 at 14pt
\font\fourteentt=cmtt12 at 14pt     \hyphenchar\fourteentt=-1
\font\fourteencsc=mtcsc10 at 14pt
\font\fourteensf=mtss10 at 14pt

\font\seventeenrm=mtr10 at 17pt
\font\seventeenbf=mtbx10 at 17pt
\font\seventeenit=mtti10 at 17pt
\font\seventeeni=mtmi10 at 17pt      \skewchar\seventeeni='177
\font\seventeensy=mtsy10 at 17pt     \skewchar\seventeensy='60
\font\seventeensl=mtsl10 at 17pt
\font\seventeentt=cmtt12 at 17pt     \hyphenchar\seventeentt=-1
\font\seventeencsc=mtcsc10 at 17pt
\font\seventeensf=mtss10 at 17pt
\else
\font\fiverm=cmr5
\font\fivei=cmmi5             \skewchar\fivei='177
\font\fivesy=cmsy5            \skewchar\fivesy='60
\font\fivebf=cmbx5

\font\sixrm=cmr6
\font\sixi=cmmi6             \skewchar\sixi='177
\font\sixsy=cmsy6            \skewchar\sixsy='60
\font\sixbf=cmbx6

\font\sevenrm=cmr7
\font\sevenit=cmti7
\font\seveni=cmmi7             \skewchar\seveni='177
\font\sevensy=cmsy7            \skewchar\sevensy='60
\font\sevenbf=cmbx7

\font\eightrm=cmr8
\font\eightbf=cmbx8
\font\eightit=cmti8
\font\eighti=cmmi8			\skewchar\eighti='177
\font\eightsy=cmsy8			\skewchar\eightsy='60
\font\eightsl=cmsl8
\font\eighttt=cmtt8			\hyphenchar\eighttt=-1
\font\eightcsc=cmcsc10 at 8pt
\font\eightsf=cmss8

\font\ninerm=cmr9
\font\ninebf=cmbx9
\font\nineit=cmti9
\font\ninei=cmmi9			\skewchar\ninei='177
\font\ninesy=cmsy9			\skewchar\ninesy='60
\font\ninesl=cmsl9
\font\ninett=cmtt9			\hyphenchar\ninett=-1
\font\ninecsc=cmcsc10 at 9pt
\font\ninesf=cmss9

\font\tenrm=cmr10
\font\tenbf=cmbx10
\font\tenit=cmti10
\font\teni=cmmi10		\skewchar\teni='177
\font\tensy=cmsy10		\skewchar\tensy='60
\font\tenex=cmex10
\font\tensl=cmsl10
\font\tentt=cmtt10		\hyphenchar\tentt=-1
\font\tencsc=cmcsc10
\font\tensf=cmss10

\font\elevenrm=cmr10 scaled \magstephalf
\font\elevenbf=cmbx10 scaled \magstephalf
\font\elevenit=cmti10 scaled \magstephalf
\font\eleveni=cmmi10 scaled \magstephalf	\skewchar\eleveni='177
\font\elevensy=cmsy10 scaled \magstephalf	\skewchar\elevensy='60
\font\elevensl=cmsl10 scaled \magstephalf
\font\eleventt=cmtt10 scaled \magstephalf	\hyphenchar\eleventt=-1
\font\elevencsc=cmcsc10 scaled \magstephalf
\font\elevensf=cmss10 scaled \magstephalf

\font\twelverm=cmr10 scaled \magstep1
\font\twelvebf=cmbx10 scaled \magstep1
\font\twelvei=cmmi10 scaled \magstep1      \skewchar\twelvei='177
\font\twelvesy=cmsy10 scaled \magstep1     \skewchar\twelvesy='60

\font\fourteenrm=cmr10 scaled \magstep2
\font\fourteenbf=cmbx10 scaled \magstep2
\font\fourteenit=cmti10 scaled \magstep2
\font\fourteeni=cmmi10 scaled \magstep2		\skewchar\fourteeni='177
\font\fourteensy=cmsy10 scaled \magstep2	\skewchar\fourteensy='60
\font\fourteensl=cmsl10 scaled \magstep2
\font\fourteentt=cmtt10 scaled \magstep2	\hyphenchar\fourteentt=-1
\font\fourteencsc=cmcsc10 scaled \magstep2
\font\fourteensf=cmss10 scaled \magstep2

\font\seventeenrm=cmr10 scaled \magstep3
\font\seventeenbf=cmbx10 scaled \magstep3
\font\seventeenit=cmti10 scaled \magstep3
\font\seventeeni=cmmi10 scaled \magstep3	\skewchar\seventeeni='177
\font\seventeensy=cmsy10 scaled \magstep3	\skewchar\seventeensy='60
\font\seventeensl=cmsl10 scaled \magstep3
\font\seventeentt=cmtt10 scaled \magstep3	\hyphenchar\seventeentt=-1
\font\seventeencsc=cmcsc10 scaled \magstep3
\font\seventeensf=cmss10 scaled \magstep3
\fi

\def\hexnumber#1{\ifcase#1 0\or1\or2\or3\or4\or5\or6\or7\or8\or9\or
  A\or B\or C\or D\or E\or F\fi}

\def\makestrut{%
  \setbox\strutbox=\hbox{%
    \vrule height.7\baselineskip depth.3\baselineskip width \z@}%
}

\def\baselinestretch{1}
\newskip\tmp@bls

\def\b@ls#1{
  \tmp@bls=#1\relax
  \baselineskip=#1\relax\makestrut
  \normalbaselineskip=\baselinestretch\tmp@bls
  \normalbaselines
}

\def\nostb@ls#1{
  \normalbaselineskip=#1\relax
  \normalbaselines
  \makestrut
}

%

\newfam\scfam  
\newfam\sffam  

\def\mit{\fam\@ne}
\def\cal{\fam\tw@}
\def\em{\ifdim\fontdimen1\font>\z@ \rm\else\it\fi}

\textfont3=\tenex
\scriptfont3=\tenex
\scriptscriptfont3=\tenex

\setbox0=\hbox{\tenex B} \p@renwd=\wd0 

\def\eightpoint{
  \def\rm{\fam0\eightrm}%
  \textfont0=\eightrm \scriptfont0=\sixrm \scriptscriptfont0=\fiverm%
  \textfont1=\eighti  \scriptfont1=\sixi  \scriptscriptfont1=\fivei%
  \textfont2=\eightsy \scriptfont2=\sixsy \scriptscriptfont2=\fivesy%
  \textfont\itfam=\eightit\def\it{\fam\itfam\eightit}%
  \ifprod@font
    \scriptfont\itfam=\sixit
      \scriptscriptfont\itfam=\fiveit
  \else
    \scriptfont\itfam=\eightit
      \scriptscriptfont\itfam=\eightit
  \fi
  \textfont\bffam=\eightbf%
    \scriptfont\bffam=\sixbf%
      \scriptscriptfont\bffam=\fivebf%
  \def\bf{\fam\bffam\eightbf}%
  \textfont\slfam=\eightsl\def\sl{\fam\slfam\eightsl}%
  \ifprod@font
    \scriptfont\slfam=\sixsl
      \scriptscriptfont\slfam=\fivesl
  \else
    \scriptfont\slfam=\eightsl
      \scriptscriptfont\slfam=\eightsl
  \fi
  \textfont\ttfam=\eighttt\def\tt{\fam\ttfam\eighttt}%
  \ifprod@font
    \scriptfont\ttfam=\sixtt
      \scriptscriptfont\ttfam=\fivett
  \else
    \scriptfont\ttfam=\eighttt
      \scriptscriptfont\ttfam=\eighttt
  \fi
  \textfont\scfam=\eightcsc\def\sc{\fam\scfam\eightcsc}%
  \ifprod@font
    \scriptfont\scfam=\sixcsc
      \scriptscriptfont\scfam=\fivecsc
  \else
    \scriptfont\scfam=\eightcsc
      \scriptscriptfont\scfam=\eightcsc
  \fi
  \textfont\sffam=\eightsf\def\sf{\fam\sffam\eightsf}%
  \ifprod@font
    \scriptfont\sffam=\sixsf
      \scriptscriptfont\sffam=\fivesf
  \else
    \scriptfont\sffam=\eightsf
      \scriptscriptfont\sffam=\eightsf
  \fi
  \def\oldstyle{\fam\@ne\eighti}%
  \b@ls{10pt}\rm\@viiipt%
}
\def\@viiipt{}

\def\ninepoint{
  \def\rm{\fam0\ninerm}%
  \textfont0=\ninerm \scriptfont0=\sixrm \scriptscriptfont0=\fiverm%
  \textfont1=\ninei  \scriptfont1=\sixi  \scriptscriptfont1=\fivei%
  \textfont2=\ninesy \scriptfont2=\sixsy \scriptscriptfont2=\fivesy%
  \textfont\itfam=\nineit\def\it{\fam\itfam\nineit}%
  \ifprod@font
    \scriptfont\itfam=\sixit
      \scriptscriptfont\itfam=\fiveit
  \else
    \scriptfont\itfam=\nineit
      \scriptscriptfont\itfam=\nineit
  \fi
  \textfont\bffam=\ninebf%
    \scriptfont\bffam=\sixbf%
      \scriptscriptfont\bffam=\fivebf%
  \def\bf{\fam\bffam\ninebf}%
  \textfont\slfam=\ninesl\def\sl{\fam\slfam\ninesl}%
  \ifprod@font
    \scriptfont\slfam=\sixsl
      \scriptscriptfont\slfam=\fivesl
  \else
    \scriptfont\slfam=\ninesl
      \scriptscriptfont\slfam=\ninesl
  \fi
  \textfont\ttfam=\ninett\def\tt{\fam\ttfam\ninett}%
  \ifprod@font
    \scriptfont\ttfam=\sixtt
      \scriptscriptfont\ttfam=\fivett
  \else
    \scriptfont\ttfam=\ninett
      \scriptscriptfont\ttfam=\ninett
  \fi
  \textfont\scfam=\ninecsc\def\sc{\fam\scfam\ninecsc}%
  \ifprod@font
    \scriptfont\scfam=\sixcsc
      \scriptscriptfont\scfam=\fivecsc
  \else
    \scriptfont\scfam=\ninecsc
      \scriptscriptfont\scfam=\ninecsc
  \fi
  \textfont\sffam=\ninesf\def\sf{\fam\sffam\ninesf}%
  \ifprod@font
    \scriptfont\sffam=\sixsf
      \scriptscriptfont\sffam=\fivesf
  \else
    \scriptfont\sffam=\ninesf
      \scriptscriptfont\sffam=\ninesf
  \fi
  \def\oldstyle{\fam\@ne\ninei}%
  \b@ls{\TextLeading plus \Feathering}\rm\@ixpt%
}
\def\@ixpt{}

\def\tenpoint{
  \def\rm{\fam0\tenrm}%
  \textfont0=\tenrm \scriptfont0=\sevenrm \scriptscriptfont0=\fiverm%
  \textfont1=\teni  \scriptfont1=\seveni  \scriptscriptfont1=\fivei%
  \textfont2=\tensy \scriptfont2=\sevensy \scriptscriptfont2=\fivesy%
  \textfont\itfam=\tenit\def\it{\fam\itfam\tenit}%
  \ifprod@font
    \scriptfont\itfam=\sevenit
      \scriptscriptfont\itfam=\fiveit
  \else
    \scriptfont\itfam=\tenit
      \scriptscriptfont\itfam=\tenit
  \fi
  \textfont\bffam=\tenbf%
    \scriptfont\bffam=\sevenbf%
      \scriptscriptfont\bffam=\fivebf%
  \def\bf{\fam\bffam\tenbf}%
  \textfont\slfam=\tensl\def\sl{\fam\slfam\tensl}%
  \ifprod@font
    \scriptfont\slfam=\sevensl
      \scriptscriptfont\slfam=\fivesl
  \else
    \scriptfont\slfam=\tensl
      \scriptscriptfont\slfam=\tensl
  \fi
  \textfont\ttfam=\tentt\def\tt{\fam\ttfam\tentt}%
  \ifprod@font
    \scriptfont\ttfam=\seventt
      \scriptscriptfont\ttfam=\fivett
  \else
    \scriptfont\ttfam=\tentt
      \scriptscriptfont\ttfam=\tentt
  \fi
  \textfont\scfam=\tencsc\def\sc{\fam\scfam\tencsc}%
  \ifprod@font
    \scriptfont\scfam=\sevencsc
      \scriptscriptfont\scfam=\fivecsc
  \else
    \scriptfont\scfam=\tencsc
      \scriptscriptfont\scfam=\tencsc
  \fi
  \textfont\sffam=\tensf\def\sf{\fam\sffam\tensf}%
  \ifprod@font
    \scriptfont\sffam=\sevensf
      \scriptscriptfont\sffam=\fivesf
  \else
    \scriptfont\sffam=\tensf
      \scriptscriptfont\sffam=\tensf
  \fi
  \def\oldstyle{\fam\@ne\teni}%
  \b@ls{11pt}\rm\@xpt%
}
\def\@xpt{}

\def\elevenpoint{
  \def\rm{\fam0\elevenrm}%
  \textfont0=\elevenrm \scriptfont0=\eightrm \scriptscriptfont0=\sixrm%
  \textfont1=\eleveni  \scriptfont1=\eighti  \scriptscriptfont1=\sixi%
  \textfont2=\elevensy \scriptfont2=\eightsy \scriptscriptfont2=\sixsy%
  \textfont\itfam=\elevenit\def\it{\fam\itfam\elevenit}%
  \ifprod@font
    \scriptfont\itfam=\eightit
      \scriptscriptfont\itfam=\sixit
  \else
    \scriptfont\itfam=\elevenit
      \scriptscriptfont\itfam=\elevenit
  \fi
  \textfont\bffam=\elevenbf%
    \scriptfont\bffam=\eightbf%
      \scriptscriptfont\bffam=\sixbf%
  \def\bf{\fam\bffam\elevenbf}%
  \textfont\slfam=\elevensl\def\sl{\fam\slfam\elevensl}%
  \ifprod@font
    \scriptfont\slfam=\eightsl
      \scriptscriptfont\slfam=\sixsl
  \else
    \scriptfont\slfam=\elevensl
      \scriptscriptfont\slfam=\elevensl
  \fi
  \textfont\ttfam=\eleventt\def\tt{\fam\ttfam\eleventt}%
  \ifprod@font
    \scriptfont\ttfam=\eighttt
      \scriptscriptfont\ttfam=\sixtt
  \else
    \scriptfont\ttfam=\eleventt
      \scriptscriptfont\ttfam=\eleventt
  \fi
  \textfont\scfam=\elevencsc\def\sc{\fam\scfam\elevencsc}%
  \ifprod@font
    \scriptfont\scfam=\eightcsc
      \scriptscriptfont\scfam=\sixcsc
  \else
    \scriptfont\scfam=\elevencsc
      \scriptscriptfont\scfam=\elevencsc
  \fi
  \textfont\sffam=\elevensf\def\sf{\fam\sffam\elevensf}%
  \ifprod@font
    \scriptfont\sffam=\eightsf
      \scriptscriptfont\sffam=\sixsf
  \else
    \scriptfont\sffam=\elevensf
      \scriptscriptfont\sffam=\elevensf
  \fi
  \def\oldstyle{\fam\@ne\eleveni}%
  \b@ls{13pt}\rm\@xipt%
}
\def\@xipt{}

\def\fourteenpoint{
  \def\rm{\fam0\fourteenrm}%
  \textfont0\fourteenrm  \scriptfont0\tenrm  \scriptscriptfont0\sevenrm%
  \textfont1\fourteeni   \scriptfont1\teni   \scriptscriptfont1\seveni%
  \textfont2\fourteensy  \scriptfont2\tensy  \scriptscriptfont2\sevensy%
  \textfont\itfam=\fourteenit\def\it{\fam\itfam\fourteenit}%
  \ifprod@font
    \scriptfont\itfam=\tenit
      \scriptscriptfont\itfam=\sevenit
  \else
    \scriptfont\itfam=\fourteenit
      \scriptscriptfont\itfam=\fourteenit
  \fi
  \textfont\bffam=\fourteenbf%
    \scriptfont\bffam=\tenbf%
      \scriptscriptfont\bffam=\sevenbf%
  \def\bf{\fam\bffam\fourteenbf}%
  \textfont\slfam=\fourteensl\def\sl{\fam\slfam\fourteensl}%
  \ifprod@font
    \scriptfont\slfam=\tensl
      \scriptscriptfont\slfam=\sevensl
  \else
    \scriptfont\slfam=\fourteensl
      \scriptscriptfont\slfam=\fourteensl
  \fi
  \textfont\ttfam=\fourteentt\def\tt{\fam\ttfam\fourteentt}%
  \ifprod@font
    \scriptfont\ttfam=\tentt
      \scriptscriptfont\ttfam=\seventt
  \else
    \scriptfont\ttfam=\fourteentt
      \scriptscriptfont\ttfam=\fourteentt
  \fi
  \textfont\scfam=\fourteencsc\def\sc{\fam\scfam\fourteencsc}%
  \ifprod@font
    \scriptfont\scfam=\tencsc
      \scriptscriptfont\scfam=\sevencsc
  \else
    \scriptfont\scfam=\fourteencsc
      \scriptscriptfont\scfam=\fourteencsc
  \fi
  \textfont\sffam=\fourteensf\def\sf{\fam\sffam\fourteensf}%
  \ifprod@font
    \scriptfont\sffam=\tensf
      \scriptscriptfont\sffam=\sevensf
  \else
    \scriptfont\sffam=\fourteensf
      \scriptscriptfont\sffam=\fourteensf
  \fi
  \def\oldstyle{\fam\@ne\fourteeni}%
  \b@ls{17pt}\rm\@xivpt%
}
\def\@xivpt{}

\def\seventeenpoint{
  \def\rm{\fam0\seventeenrm}%
  \textfont0\seventeenrm  \scriptfont0\twelverm  \scriptscriptfont0\tenrm%
  \textfont1\seventeeni   \scriptfont1\twelvei   \scriptscriptfont1\teni%
  \textfont2\seventeensy  \scriptfont2\twelvesy  \scriptscriptfont2\tensy%
  \textfont\itfam=\seventeenit\def\it{\fam\itfam\seventeenit}%
  \ifprod@font
    \scriptfont\itfam=\twelveit
      \scriptscriptfont\itfam=\tenit
  \else
    \scriptfont\itfam=\seventeenit
      \scriptscriptfont\itfam=\seventeenit
  \fi
  \textfont\bffam=\seventeenbf%
    \scriptfont\bffam=\twelvebf%
      \scriptscriptfont\bffam=\tenbf%
  \def\bf{\fam\bffam\seventeenbf}%
  \textfont\slfam=\seventeensl\def\sl{\fam\slfam\seventeensl}%
  \ifprod@font
    \scriptfont\slfam=\twelvesl
      \scriptscriptfont\slfam=\tensl
  \else
    \scriptfont\slfam=\seventeensl
      \scriptscriptfont\slfam=\seventeensl
  \fi
  \textfont\ttfam=\seventeentt\def\tt{\fam\ttfam\seventeentt}%
  \ifprod@font
    \scriptfont\ttfam=\twelvett
      \scriptscriptfont\ttfam=\tentt
  \else
    \scriptfont\ttfam=\seventeentt
      \scriptscriptfont\ttfam=\seventeentt
  \fi
  \textfont\scfam=\seventeencsc\def\sc{\fam\scfam\seventeencsc}%
  \ifprod@font
    \scriptfont\scfam=\twelvecsc
      \scriptscriptfont\scfam=\tencsc
  \else
    \scriptfont\scfam=\seventeencsc
      \scriptscriptfont\scfam=\seventeencsc
  \fi
  \textfont\sffam=\seventeensf\def\sf{\fam\sffam\seventeensf}%
  \ifprod@font
    \scriptfont\sffam=\twelvesf
      \scriptscriptfont\sffam=\tensf
  \else
    \scriptfont\sffam=\seventeensf
      \scriptscriptfont\sffam=\seventeensf
  \fi
  \def\oldstyle{\fam\@ne\seventeeni}%
  \b@ls{20pt}\rm\@xviipt%
}
\def\@xviipt{}

\lineskip=1pt      \normallineskip=\lineskip
\lineskiplimit=\z@ \normallineskiplimit=\lineskiplimit


\def\loadboldmathnames{%
  \def\balpha{{\bmath{\alpha}}}%
  \def\bbeta{{\bmath{\beta}}}%
  \def\bgamma{{\bmath{\gamma}}}%
  \def\bdelta{{\bmath{\delta}}}%
  \def\bepsilon{{\bmath{\epsilon}}}%
  \def\bzeta{{\bmath{\zeta}}}%
  \def\boldeta{{\bmath{\eta}}}%
  \def\btheta{{\bmath{\theta}}}%
  \def\biota{{\bmath{\iota}}}%
  \def\bkappa{{\bmath{\kappa}}}%
  \def\blambda{{\bmath{\lambda}}}%
  \def\bmu{{\bmath{\mu}}}%
  \def\bnu{{\bmath{\nu}}}%
  \def\bxi{{\bmath{\xi}}}%
  \def\bpi{{\bmath{\pi}}}%
  \def\brho{{\bmath{\rho}}}%
  \def\bsigma{{\bmath{\sigma}}}%
  \def\btau{{\bmath{\tau}}}%
  \def\bupsilon{{\bmath{\upsilon}}}%
  \def\bphi{{\bmath{\phi}}}%
  \def\bchi{{\bmath{\chi}}}%
  \def\bpsi{{\bmath{\psi}}}%
  \def\bomega{{\bmath{\omega}}}%
  \def\bvarepsilon{{\bmath{\varepsilon}}}%
  \def\bvartheta{{\bmath{\vartheta}}}%
  \def\bvarpi{{\bmath{\varpi}}}%
  \def\bvarrho{{\bmath{\varrho}}}%
  \def\bvarsigma{{\bmath{\varsigma}}}%
  \def\bvarphi{{\bmath{\varphi}}}%
  \def\baleph{{\bmath{\aleph}}}%
  \def\bimath{{\bmath{\imath}}}%
  \def\bjmath{{\bmath{\jmath}}}%
  \def\bell{{\bmath{\ell}}}%
  \def\bwp{{\bmath{\wp}}}%
  \def\bRe{{\bmath{\Re}}}%
  \def\bIm{{\bmath{\Im}}}%
  \def\bpartial{{\bmath{\partial}}}%
  \def\binfty{{\bmath{\infty}}}%
  \def\bprime{{\bmath{\prime}}}%
  \def\bemptyset{{\bmath{\emptyset}}}%
  \def\bnabla{{\bmath{\nabla}}}%
  \def\btop{{\bmath{\top}}}%
  \def\bbot{{\bmath{\bot}}}%
  \def\btriangle{{\bmath{\triangle}}}%
  \def\bforall{{\bmath{\forall}}}%
  \def\bexists{{\bmath{\exists}}}%
  \def\bneg{{\bmath{\neg}}}%
  \def\bflat{{\bmath{\flat}}}%
  \def\bnatural{{\bmath{\natural}}}%
  \def\bsharp{{\bmath{\sharp}}}%
  \def\bclubsuit{{\bmath{\clubsuit}}}%
  \def\bdiamondsuit{{\bmath{\diamondsuit}}}%
  \def\bheartsuit{{\bmath{\heartsuit}}}%
  \def\bspadesuit{{\bmath{\spadesuit}}}%
  \def\bsmallint{{\bmath{\smallint}}}%
  \def\btriangleleft{{\bmath{\triangleleft}}}%
  \def\btriangleright{{\bmath{\triangleright}}}%
  \def\bbigtriangleup{{\bmath{\bigtriangleup}}}%
  \def\bbigtriangledown{{\bmath{\bigtriangledown}}}%
  \def\bwedge{{\bmath{\wedge}}}%
  \def\bvee{{\bmath{\vee}}}%
  \def\bcap{{\bmath{\cap}}}%
  \def\bcup{{\bmath{\cup}}}%
  \def\bddagger{{\bmath{\ddagger}}}%
  \def\bdagger{{\bmath{\dagger}}}%
  \def\bsqcap{{\bmath{\sqcap}}}%
  \def\bsqcup{{\bmath{\sqcup}}}%
  \def\buplus{{\bmath{\uplus}}}%
  \def\bamalg{{\bmath{\amalg}}}%
  \def\bdiamond{{\bmath{\diamond}}}%
  \def\bbullet{{\bmath{\bullet}}}%
  \def\bwr{{\bmath{\wr}}}%
  \def\bdiv{{\bmath{\div}}}%
  \def\bodot{{\bmath{\odot}}}%
  \def\boslash{{\bmath{\oslash}}}%
  \def\botimes{{\bmath{\otimes}}}%
  \def\bominus{{\bmath{\ominus}}}%
  \def\boplus{{\bmath{\oplus}}}%
  \def\bmp{{\bmath{\mp}}}%
  \def\bpm{{\bmath{\pm}}}%
  \def\bcirc{{\bmath{\circ}}}%
  \def\bbigcirc{{\bmath{\bigcirc}}}%
  \def\bsetminus{{\bmath{\setminus}}}%
  \def\bcdot{{\bmath{\cdot}}}%
  \def\bast{{\bmath{\ast}}}%
  \def\btimes{{\bmath{\times}}}%
  \def\bstar{{\bmath{\star}}}%
  \def\bpropto{{\bmath{\propto}}}%
  \def\bsqsubseteq{{\bmath{\sqsubseteq}}}%
  \def\bsqsupseteq{{\bmath{\sqsupseteq}}}%
  \def\bparallel{{\bmath{\parallel}}}%
  \def\bmid{{\bmath{\mid}}}%
  \def\bdashv{{\bmath{\dashv}}}%
  \def\bvdash{{\bmath{\vdash}}}%
  \def\bnearrow{{\bmath{\nearrow}}}%
  \def\bsearrow{{\bmath{\searrow}}}%
  \def\bnwarrow{{\bmath{\nwarrow}}}%
  \def\bswarrow{{\bmath{\swarrow}}}%
  \def\bLeftrightarrow{{\bmath{\Leftrightarrow}}}%
  \def\bLeftarrow{{\bmath{\Leftarrow}}}%
  \def\bRightarrow{{\bmath{\Rightarrow}}}%
  \def\bleq{{\bmath{\leq}}}%
  \def\bgeq{{\bmath{\geq}}}%
  \def\bsucc{{\bmath{\succ}}}%
  \def\bprec{{\bmath{\prec}}}%
  \def\bapprox{{\bmath{\approx}}}%
  \def\bsucceq{{\bmath{\succeq}}}%
  \def\bpreceq{{\bmath{\preceq}}}%
  \def\bsupset{{\bmath{\supset}}}%
  \def\bsubset{{\bmath{\subset}}}%
  \def\bsupseteq{{\bmath{\supseteq}}}%
  \def\bsubseteq{{\bmath{\subseteq}}}%
  \def\bin{{\bmath{\in}}}%
  \def\bni{{\bmath{\ni}}}%
  \def\bgg{{\bmath{\gg}}}%
  \def\bll{{\bmath{\ll}}}%
  \def\bnot{{\bmath{\not}}}%
  \def\bleftrightarrow{{\bmath{\leftrightarrow}}}%
  \def\bleftarrow{{\bmath{\leftarrow}}}%
  \def\brightarrow{{\bmath{\rightarrow}}}%
  \def\bmapstochar{{\bmath{\mapstochar}}}%
  \def\bsim{{\bmath{\sim}}}%
  \def\bsimeq{{\bmath{\simeq}}}%
  \def\bperp{{\bmath{\perp}}}%
  \def\bequiv{{\bmath{\equiv}}}%
  \def\basymp{{\bmath{\asymp}}}%
  \def\bsmile{{\bmath{\smile}}}%
  \def\bfrown{{\bmath{\frown}}}%
  \def\bleftharpoonup{{\bmath{\leftharpoonup}}}%
  \def\bleftharpoondown{{\bmath{\leftharpoondown}}}%
  \def\brightharpoonup{{\bmath{\rightharpoonup}}}%
  \def\brightharpoondown{{\bmath{\rightharpoondown}}}%
  \def\blhook{{\bmath{\lhook}}}%
  \def\brhook{{\bmath{\rhook}}}%
  \def\bldotp{{\bmath{\ldotp}}}%
  \def\bcdotp{{\bmath{\cdotp}}}%
}

\def\,{\relax\ifmmode \mskip\thinmuskip\else \thinspace\fi}
\let\protect=\relax

\long\def\@ifundefined#1#2#3{\expandafter\ifx\csname
  #1\endcsname\relax#2\else#3\fi}




\newtoks\math@groups \math@groups={}
\def\addtom@thgroup#1#2{#1\expandafter{\the#1#2}} 



\def\addtosizeh@ok#1#2#3#4{%
  \expandafter\def\csname @#1pt\endcsname{%
    \def\s@ze{#2}\def\ss@ze{#3}\def\sss@ze{#4}\the\math@groups%
  }%
}



\let\resetsizehook=\addtosizeh@ok


\ifprod@font
  \addtosizeh@ok{viii} {8} {6}  {5}
  \addtosizeh@ok{ix}   {9} {6}  {5}
  \addtosizeh@ok{x}    {10}{7}  {5}
  \addtosizeh@ok{xi}   {11}{8}  {6}
  \addtosizeh@ok{xiv}  {14}{10} {7}
  \addtosizeh@ok{xvii} {17}{12}{10}
\else
  \addtosizeh@ok{viii} {8}     {6}     {5}
  \addtosizeh@ok{ix}   {9}     {6}     {5}
  \addtosizeh@ok{x}    {10}    {7}     {5}
  \addtosizeh@ok{xi}   {10.95} {8}     {6}
  \addtosizeh@ok{xiv}  {14.4}  {10}    {7}
  \addtosizeh@ok{xvii} {17.28} {12}    {10}
\fi

\def\get@font#1#2#3{%
  \edef\fonts@ze{\romannumeral#3}
  \edef\fontn@me{\fonts@ze#1}
  \@ifundefined{\fontn@me}%
    {
     \global\expandafter\font\csname \fontn@me\endcsname=#2 at #3pt}%
    {}%
}

\def\ass@tfont#1#2{%
  \xdef\fam@name{\csname #1\endcsname}%
  \xdef\font@name{\csname #2\endcsname}%
  \let\textfont@name\font@name
  \textfont\fam@name\textfont@name
}

\def\ass@sfont#1#2{%
  \xdef\fam@name{\csname #1\endcsname}%
  \xdef\font@name{\csname #2\endcsname}%
  \let\textfont@name\font@name
  \scriptfont\fam@name\textfont@name
}

\def\ass@ssfont#1#2{%
  \xdef\fam@name{\csname #1\endcsname}%
  \xdef\font@name{\csname #2\endcsname}%
  \let\textfont@name\font@name
  \scriptscriptfont\fam@name\textfont@name
}


\def\NewSymbolFont#1#2{%
  \expandafter\ifx\csname sym#1fam\endcsname\relax 
    \expandafter\newfam\csname sym#1fam\endcsname
    \expandafter\edef\csname sym#1fam\endcsname{\the\allocationnumber}%
    \addtom@thgroup\math@groups{%
      \get@font{#1}{#2}{\s@ze}%
      \ass@tfont{sym#1fam}{\fontn@me}%
      \get@font{#1}{#2}{\ss@ze}%
      \ass@sfont{sym#1fam}{\fontn@me}%
      \get@font{#1}{#2}{\sss@ze}%
      \ass@ssfont{sym#1fam}{\fontn@me}%
    }%
  \else
    \errmessage{Family `#1' already defined}%
  \fi
}


\def\NewMathSymbol#1#2#3#4{%
  \edef\f@mly{\expandafter\hexnumber{\csname sym#3fam\endcsname}}%
  \mathchardef#1="#2\f@mly#4\relax
}


\newif\ifd@f

\def\NewMathDelimiter#1#2#3#4#5#6{%
  \d@ftrue
  \expandafter\ifx\csname sym#3fam\endcsname\relax
    \d@ffalse \errmessage{Family `#3' is not defined}%
  \fi
  \expandafter\ifx\csname sym#5fam\endcsname\relax
    \d@ffalse \errmessage{Family `#5' is not defined}%
  \fi
  \ifd@f
    \edef\f@mly{\expandafter\hexnumber{\csname sym#3fam\endcsname}}%
    \edef\f@mlytw@{\expandafter\hexnumber{\csname sym#5fam\endcsname}}%
    \xdef#1{\delimiter"#2\f@mly #4\f@mlytw@ #6\relax}%
  \fi
}


\def\setboxz@h{\setbox\z@\hbox}
\def\wdz@{\wd\z@}
\def\boxz@{\box\z@}
\def\setbox@ne{\setbox\@ne}
\def\wd@ne{\wd\@ne}

\def\math@atom#1#2{%
   \binrel@{#1}\binrel@@{#2}}
\def\binrel@#1{\setboxz@h{\thinmuskip0mu
  \medmuskip\m@ne mu\thickmuskip\@ne mu$#1\m@th$}%
 \setbox@ne\hbox{\thinmuskip0mu\medmuskip\m@ne mu\thickmuskip
  \@ne mu${}#1{}\m@th$}%
 \setbox\tw@\hbox{\hskip\wd@ne\hskip-\wdz@}}
\def\binrel@@#1{\ifdim\wd2<\z@\mathbin{#1}\else\ifdim\wd\tw@>\z@
 \mathrel{#1}\else{#1}\fi\fi}

\def\m@thit{1}

\def\set@skchar#1{\global\expandafter\skewchar
  \csname\fontn@me\endcsname=#1\relax}

\def\NewMathAlphabet#1#2#3{%
  \def\tst{#3}%
  \ifx\tst\empty\else 
    \expandafter\gdef\csname #1@sc\endcsname{}
  \fi
  \expandafter\def\csname #1\endcsname{
    \protect\csname @#1\endcsname}%
  \expandafter\def\csname @#1\endcsname##1{
    {%
    \begingroup
      \get@font{#1}{#2}{\s@ze}%
      \@ifundefined{#1@sc}{}{\set@skchar{#3}}%
      \ass@tfont{m@thit}{\fontn@me}%
      \get@font{#1}{#2}{\ss@ze}%
      \@ifundefined{#1@sc}{}{\set@skchar{#3}}%
      \ass@sfont{m@thit}{\fontn@me}%
      \get@font{#1}{#2}{\sss@ze}%
      \@ifundefined{#1@sc}{}{\set@skchar{#3}}%
      \ass@ssfont{m@thit}{\fontn@me}%
      \math@atom{##1}{%
      \mathchoice%
        {\hbox{$\m@th\displaystyle##1$}}%
        {\hbox{$\m@th\textstyle##1$}}%
        {\hbox{$\m@th\scriptstyle##1$}}%
        {\hbox{$\m@th\scriptscriptstyle##1$}}}%
    \endgroup
    }%
  }%
}


\newif\iffirstta  \firsttatrue

\def\set@hchar#1{\global\expandafter\hyphenchar
  \csname\fontn@me\endcsname=#1\relax}

\def\NewTextAlphabet#1#2#3{%
  \iffirstta
    \global\firsttafalse
    \newfam\scratchfam
    \edef\scrt@fam{\the\allocationnumber}%
  \fi
  \def\tst{#3}%
  \ifx\tst\empty\else 
    \expandafter\gdef\csname #1@hc\endcsname{}
  \fi
  \expandafter\def\csname #1\endcsname{
    \protect\csname t@#1\endcsname}%
  \long\expandafter\def\csname t@#1\endcsname##1{
    \ifmmode
      \typeout{Warning: do not use \expandafter\string\csname #1\endcsname
        \space in math mode}\fi%
    {%
      \get@font{#1}{#2}{\s@ze}\let\t@xtfnt=\fontn@me\relax
      \@ifundefined{#1@hc}{}{\set@hchar{#3}}%
      \ass@tfont{scrt@fam}{\fontn@me}%
      \get@font{#1}{#2}{\ss@ze}%
      \@ifundefined{#1@hc}{}{\set@hchar{#3}}%
      \ass@sfont{scrt@fam}{\fontn@me}%
      \get@font{#1}{#2}{\sss@ze}%
      \@ifundefined{#1@hc}{}{\set@hchar{#3}}%
      \ass@ssfont{scrt@fam}{\fontn@me}%
      \fam\scratchfam\csname\t@xtfnt\endcsname
    ##1%
    }%
  }%
  \expandafter\def\csname #1shape
    \endcsname{\protect\csname @#1shape\endcsname}%
  \expandafter\def\csname @#1shape\endcsname{
    \ifmmode
      \typeout{Warning: do not use \expandafter\string\csname
        #1shape\endcsname \space in math mode}\fi
      \get@font{#1}{#2}{\s@ze}\let\t@xtfnt=\fontn@me\relax
      \@ifundefined{#1@hc}{}{\set@hchar{#3}}%
      \ass@tfont{scrt@fam}{\fontn@me}%
      \get@font{#1}{#2}{\ss@ze}%
      \@ifundefined{#1@hc}{}{\set@hchar{#3}}%
      \ass@sfont{scrt@fam}{\fontn@me}%
      \get@font{#1}{#2}{\sss@ze}%
      \@ifundefined{#1@hc}{}{\set@hchar{#3}}%
      \ass@ssfont{scrt@fam}{\fontn@me}%
      \fam\scratchfam\csname\t@xtfnt\endcsname
  }%
}


\ifprod@font
  \def\math@itfnt{mtmib10}
  \def\math@syfnt{mtbsy10}
\else
  \def\math@itfnt{cmmib10}
  \def\math@syfnt{cmbsy10}
\fi

\def\m@thsy{2}

\def\bmath{\protect\@bmath}
\def\@bmath#1{%
  {%
  \begingroup
    \get@font{mthit}{\math@itfnt}{\s@ze}\set@skchar{'177}%
    \ass@tfont{m@thit}{\fontn@me}%
    \get@font{mthit}{\math@itfnt}{\ss@ze}\set@skchar{'177}%
    \ass@sfont{m@thit}{\fontn@me}%
    \get@font{mthit}{\math@itfnt}{\sss@ze}\set@skchar{'177}%
    \ass@ssfont{m@thit}{\fontn@me}%
    \get@font{mthsy}{\math@syfnt}{\s@ze}\set@skchar{'60}%
    \ass@tfont{m@thsy}{\fontn@me}%
    \get@font{mthsy}{\math@syfnt}{\ss@ze}\set@skchar{'60}%
    \ass@sfont{m@thsy}{\fontn@me}%
    \get@font{mthsy}{\math@syfnt}{\sss@ze}\set@skchar{'60}%
    \ass@ssfont{m@thsy}{\fontn@me}%
    \math@atom{#1}{%
    \mathchoice%
      {\hbox{$\m@th\displaystyle#1$}}%
      {\hbox{$\m@th\textstyle#1$}}%
      {\hbox{$\m@th\scriptstyle#1$}}%
      {\hbox{$\m@th\scriptscriptstyle#1$}}}%
  \endgroup
  }%
}



\def\la{\mathrel{\mathchoice {\vcenter{\offinterlineskip\halign{\hfil
$\displaystyle##$\hfil\cr<\cr\sim\cr}}}
{\vcenter{\offinterlineskip\halign{\hfil$\textstyle##$\hfil\cr
<\cr\sim\cr}}}
{\vcenter{\offinterlineskip\halign{\hfil$\scriptstyle##$\hfil\cr
<\cr\sim\cr}}}
{\vcenter{\offinterlineskip\halign{\hfil$\scriptscriptstyle##$\hfil\cr
<\cr\sim\cr}}}}}

\def\diameter{{\ifmmode\mathchoice
{\ooalign{\hfil\hbox{$\displaystyle/$}\hfil\crcr
{\hbox{$\displaystyle\mathchar"20D$}}}}
{\ooalign{\hfil\hbox{$\textstyle/$}\hfil\crcr
{\hbox{$\textstyle\mathchar"20D$}}}}
{\ooalign{\hfil\hbox{$\scriptstyle/$}\hfil\crcr
{\hbox{$\scriptstyle\mathchar"20D$}}}}
{\ooalign{\hfil\hbox{$\scriptscriptstyle/$}\hfil\crcr
{\hbox{$\scriptscriptstyle\mathchar"20D$}}}}
\else{\ooalign{\hfil/\hfil\crcr\mathhexbox20D}}%
\fi}}

\def\sq{\ifmmode\squareforqed\else{\unskip\nobreak\hfil
\penalty50\hskip1em\null\nobreak\hfil\squareforqed
\parfillskip=0pt\finalhyphendemerits=0\endgraf}\fi}
\def\squareforqed{\hbox{\rlap{$\sqcap$}$\sqcup$}}

\def\fs{\hbox{$.\!\!^{\rm s}$}}

\def\arcsec{\hbox{$^{\prime\prime}$}}


\def\bbbc{{\mathchoice {\setbox0=\hbox{$\displaystyle\rm C$}\hbox{\hbox
to0pt{\kern0.4\wd0\vrule height0.9\ht0\hss}\box0}}
{\setbox0=\hbox{$\textstyle\rm C$}\hbox{\hbox
to0pt{\kern0.4\wd0\vrule height0.9\ht0\hss}\box0}}
{\setbox0=\hbox{$\scriptstyle\rm C$}\hbox{\hbox
to0pt{\kern0.4\wd0\vrule height0.9\ht0\hss}\box0}}
{\setbox0=\hbox{$\scriptscriptstyle\rm C$}\hbox{\hbox
to0pt{\kern0.4\wd0\vrule height0.9\ht0\hss}\box0}}}}
\def\bbbq{{\mathchoice {\setbox0=\hbox{$\displaystyle\rm
Q$}\hbox{\raise
0.15\ht0\hbox to0pt{\kern0.4\wd0\vrule height0.8\ht0\hss}\box0}}
{\setbox0=\hbox{$\textstyle\rm Q$}\hbox{\raise
0.15\ht0\hbox to0pt{\kern0.4\wd0\vrule height0.8\ht0\hss}\box0}}
{\setbox0=\hbox{$\scriptstyle\rm Q$}\hbox{\raise
0.15\ht0\hbox to0pt{\kern0.4\wd0\vrule height0.7\ht0\hss}\box0}}
{\setbox0=\hbox{$\scriptscriptstyle\rm Q$}\hbox{\raise
0.15\ht0\hbox to0pt{\kern0.4\wd0\vrule height0.7\ht0\hss}\box0}}}}
\def\bbbt{{\mathchoice {\setbox0=\hbox{$\displaystyle\rm
T$}\hbox{\hbox to0pt{\kern0.3\wd0\vrule height0.9\ht0\hss}\box0}}
{\setbox0=\hbox{$\textstyle\rm T$}\hbox{\hbox
to0pt{\kern0.3\wd0\vrule height0.9\ht0\hss}\box0}}
{\setbox0=\hbox{$\scriptstyle\rm T$}\hbox{\hbox
to0pt{\kern0.3\wd0\vrule height0.9\ht0\hss}\box0}}
{\setbox0=\hbox{$\scriptscriptstyle\rm T$}\hbox{\hbox
to0pt{\kern0.3\wd0\vrule height0.9\ht0\hss}\box0}}}}
\def\bbbs{{\mathchoice
{\setbox0=\hbox{$\displaystyle     \rm S$}\hbox{\raise0.5\ht0\hbox
to0pt{\kern0.35\wd0\vrule height0.45\ht0\hss}\hbox
to0pt{\kern0.55\wd0\vrule height0.5\ht0\hss}\box0}}
{\setbox0=\hbox{$\textstyle        \rm S$}\hbox{\raise0.5\ht0\hbox
to0pt{\kern0.35\wd0\vrule height0.45\ht0\hss}\hbox
to0pt{\kern0.55\wd0\vrule height0.5\ht0\hss}\box0}}
{\setbox0=\hbox{$\scriptstyle      \rm S$}\hbox{\raise0.5\ht0\hbox
to0pt{\kern0.35\wd0\vrule height0.45\ht0\hss}\raise0.05\ht0\hbox
to0pt{\kern0.5\wd0\vrule height0.45\ht0\hss}\box0}}
{\setbox0=\hbox{$\scriptscriptstyle\rm S$}\hbox{\raise0.5\ht0\hbox
to0pt{\kern0.4\wd0\vrule height0.45\ht0\hss}\raise0.05\ht0\hbox
to0pt{\kern0.55\wd0\vrule height0.45\ht0\hss}\box0}}}}
\def\bbbz{{\mathchoice {\hbox{$\sf\textstyle Z\kern-0.4em Z$}}
{\hbox{$\sf\textstyle Z\kern-0.4em Z$}}
{\hbox{$\sf\scriptstyle Z\kern-0.3em Z$}}
{\hbox{$\sf\scriptscriptstyle Z\kern-0.2em Z$}}}}


\def\Nulle{0} 
\def\Afe{1}   
\def\Hae{2}   
\def\Hbe{3}   
\def\Hce{4}   
\def\Hde{5}   


\newcount\LastMac       \LastMac=\Nulle

\newskip\half      \half=5.5pt plus 1.5pt minus 2.25pt
\newskip\one       \one=11pt plus 3pt minus 5.5pt
\newskip\onehalf   \onehalf=16.5pt plus 5.5pt minus 8.25pt
\newskip\two       \two=22pt plus 5.5pt minus 11pt

\def\Half{\addvspace{\half}}
\def\One{\addvspace{\one}}
\def\OneHalf{\addvspace{\onehalf}}
\def\Two{\addvspace{\two}}

\def\Raggedright{
  \rightskip=\z@ plus \hsize\relax
}

\def\Fullout{
  \rightskip=\z@\relax
}

\def\Hang#1#2{
  \hangindent=#1%
  \hangafter=#2\relax
}


\newif\ifsp@page
\def\pagestyle#1{\csname ps@#1\endcsname}
\def\thispagestyle#1{\global\sp@pagetrue\gdef\sp@type{#1}}

\def\ps@titlepage{%
  \def\@oddhead{\eightpoint\noindent \the\CatchLine
    \ifprod@font\else\qquad Printed\ \today\qquad
      (MN plain \TeX\ macros\ v\@version)\fi \hfil}%
  \let\@evenhead=\@oddhead
  \def\@oddfoot{\eightpoint\copyright\ \@pubyear\ RAS\hfil}%
  \def\@evenfoot{\hfil\eightpoint\noindent\copyright\ \@pubyear\ RAS}%
}

\def\ps@headings{%
  \def\@oddhead{\elevenpoint\it\noindent
    \hfill\the\RightHeader\hskip1.5em\rm\folio}%
  \def\@evenhead{\elevenpoint\noindent
    \folio\hskip1.5em\it\the\LeftHeader\hfill}%
  \def\@oddfoot{\eightpoint\noindent\copyright\ \@pubyear\ RAS,
    MNRAS {\bf \@volume}, \@pagerange\hfil}%
  \def\@evenfoot{\hfil\eightpoint\copyright\ \@pubyear\ RAS,
    MNRAS {\bf \@volume}, \@pagerange}%
}

\def\ps@plate{%
  \def\@oddhead{\eightpoint\noindent\plt@cap\hfil}%
  \def\@evenhead{\eightpoint\noindent\plt@cap\hfil}%
  \def\@oddfoot{\eightpoint\noindent\copyright\ \@pubyear\ RAS,
    MNRAS {\bf \@volume}, \@pagerange\hfil}%
  \def\@evenfoot{\hfil\eightpoint\copyright\ \@pubyear\ RAS,
    MNRAS {\bf \@volume}, \@pagerange}%
}



\def\title#1{
  \bgroup
    \vbox to 8pt{\vss}%
    \seventeenpoint
    \Raggedright
    \noindent \strut{\bf #1}\par
  \egroup
}

\def\author#1{
  \bgroup
    \ifnum\LastMac=\Afe \OneHalf\else \vskip 21pt\fi
    \fourteenpoint
    \Raggedright
    \noindent \strut #1\par
    \vskip 3pt%
  \egroup
}

\def\affiliation#1{
  \bgroup
    \vskip -4pt%
    \eightpoint
    \Raggedright
    \noindent \strut {\it #1}\par
  \egroup
  \LastMac=\Afe\relax
}

\def\acceptedline#1{
  \bgroup
    \Two
    \eightpoint
    \Raggedright
    \noindent \strut #1\par
  \egroup
}

\long\def\abstract#1{%
  \bgroup
    \vskip 20pt%
    \leftskip 11pc\rightskip\z@
    \noindent{\ninebf ABSTRACT}\par
    \tenpoint
    \Fullout
    \noindent #1\par
  \egroup
}

\long\def\keywords#1{
  \bgroup
    \Half
    \leftskip 11pc\rightskip\z@
    \tenpoint
    \Fullout
    \noindent\hbox{\bf Key words:}\ #1\par
  \egroup
}


\def\maketitle{%
  \EndOpening
  \ifsinglecol \else \MakePage\fi
}


\def\pageoffset#1#2{\hoffset=#1\relax\voffset=#2\relax}


\def\@nameuse#1{\csname #1\endcsname}
\def\arabic#1{\@arabic{\@nameuse{#1}}}
\def\alph#1{\@alph{\@nameuse{#1}}}
\def\Alph#1{\@Alph{\@nameuse{#1}}}
\def\@arabic#1{\number #1}
\def\@Alph#1{\ifcase#1\or A\or B\or C\or D\else\@Ialph{#1}\fi}
\def\@Ialph#1{\ifcase#1\or \or \or \or \or E\or F\or G\or H\or I\or J\or
   K\or L\or M\or N\or O\or P\or Q\or R\or S\or T\or U\or V\or W\or X\or
   Y\or Z\else\errmessage{Counter out of range}\fi}
\def\@alph#1{\ifcase#1\or a\or b\or c\or d\else\@ialph{#1}\fi}
\def\@ialph#1{\ifcase#1\or \or \or \or \or e\or f\or g\or h\or i\or j\or
   k\or l\or m\or n\or o\or p\or q\or r\or s\or t\or u\or v\or w\or x\or y\or
   z\else\errmessage{Counter out of range}\fi}


\newcount\Eqnno
\newcount\SubEqnno

\def\theeq{\arabic{Eqnno}}
\def\thesubeq{\alph{SubEqnno}}

\def\stepeq{\relax
  \global\SubEqnno \z@
  \global\advance\Eqnno \@ne\relax
  {\rm (\theeq)}%
}

\def\startsubeq{\relax
  \global\SubEqnno \z@
  \global\advance\Eqnno \@ne\relax
  \stepsubeq
}

\def\stepsubeq{\relax
  \global\advance\SubEqnno \@ne\relax
  {\rm (\theeq\thesubeq)}%
}


\newcount\Sec        
\newcount\SecSec
\newcount\SecSecSec

\def\thesection{\arabic{Sec}}
\def\thesubsection{\thesection.\arabic{SecSec}}
\def\thesubsubsection{\thesubsection.\arabic{SecSecSec}}

\Sec=\z@

\def\:{\let\@sptoken= } \:  
\def\:{\@xifnch} \expandafter\def\: {\futurelet\@tempc\@ifnch}

\def\@ifnextchar#1#2#3{%
  \let\@tempMACe #1%
  \def\@tempMACa{#2}%
  \def\@tempMACb{#3}%
  \futurelet \@tempMACc\@ifnch%
}

\def\@ifnch{%
\ifx \@tempMACc \@sptoken%
  \let\@tempMACd\@xifnch%
\else%
  \ifx \@tempMACc \@tempMACe%
    \let\@tempMACd\@tempMACa%
  \else%
    \let\@tempMACd\@tempMACb%
  \fi%
\fi%
\@tempMACd%
}

\def\@ifstar#1#2{\@ifnextchar *{\def\@tempMACa*{#1}\@tempMACa}{#2}}

\newskip\@tempskipb

\def\addvspace#1{%
  \ifvmode\else \endgraf\fi%
  \ifdim\lastskip=\z@%
    \vskip #1\relax%
  \else%
    \@tempskipb#1\relax\@xaddvskip%
  \fi%
}

\def\@xaddvskip{%
  \ifdim\lastskip<\@tempskipb%
    \vskip-\lastskip%
    \vskip\@tempskipb\relax%
  \else%
    \ifdim\@tempskipb<\z@%
      \ifdim\lastskip<\z@ \else%
        \advance\@tempskipb\lastskip%
        \vskip-\lastskip\vskip\@tempskipb%
      \fi%
    \fi%
  \fi%
}

\newskip\@tmpSKIP

\def\addpen#1{%
  \ifvmode
    \if@nobreak
    \else
      \ifdim\lastskip=\z@
        \penalty#1\relax
      \else
        \@tmpSKIP=\lastskip
        \vskip -\lastskip
        \penalty#1\vskip\@tmpSKIP
      \fi
    \fi
  \fi
}

\newcount\@clubpen   \@clubpen=\clubpenalty
\newif\if@nobreak    \@nobreakfalse

\def\@noafterindent{%
  \global\@nobreaktrue
  \everypar{\if@nobreak
              \global\@nobreakfalse
              \clubpenalty \@M
              {\setbox\z@\lastbox}%
              \LastMac=\Nulle\relax%
            \else
              \clubpenalty \@clubpen
              \everypar{}%
            \fi}%
}

\newcount\gds@cbrk   \gds@cbrk=-300

\def\@nohdbrk{\interlinepenalty \@M\relax}

\let\@par=\par
\def\@restorepar{\def\par{\@par}}

\newif\if@endpe   \@endpefalse
 
\def\@doendpe{\@endpetrue \@nobreakfalse \LastMac=\Nulle\relax%
     \def\par{\@restorepar\everypar{}\par\@endpefalse}%
              \everypar{\setbox\z@\lastbox\everypar{}\@endpefalse}%
}

\def\section{\@ifstar{\@ssection}{\@section}}

\def\@section#1{
  \if@nobreak
    \everypar{}%
    \ifnum\LastMac=\Hae \addvspace{\half}\fi
  \else
    \addpen{\gds@cbrk}%
    \addvspace{\two}%
  \fi
  \bgroup
    \ninepoint\bf
    \Raggedright
    \global\advance\Sec \@ne
    \ifappendix
      \global\Eqnno=\z@ \global\SubEqnno=\z@\relax
      \def\ch@ck{#1}%
      \ifx\ch@ck\empty \def\c@lon{}\else\def\c@lon{:}\fi
      \noindent\@nohdbrk APPENDIX\ \thesection\c@lon\hskip 0.5em%
        \uppercase{#1}\par
    \else
      \noindent\@nohdbrk\thesection\hskip 1pc \uppercase{#1}\par
    \fi
    \global\SecSec=\z@
  \egroup
  \nobreak
  \vskip\half
  \nobreak
  \@noafterindent
  \LastMac=\Hae\relax
}

\def\@ssection#1{
  \if@nobreak
    \everypar{}%
    \ifnum\LastMac=\Hae \addvspace{\half}\fi
  \else
    \addpen{\gds@cbrk}%
    \addvspace{\two}%
  \fi
  \bgroup
    \ninepoint\bf
    \Raggedright
    \noindent\@nohdbrk\uppercase{#1}\par
  \egroup
  \nobreak
  \vskip\half
  \nobreak
  \@noafterindent
  \LastMac=\Hae\relax
}

\def\subsection{\@ifstar{\@ssubsection}{\@subsection}}

\def\@subsection#1{
  \if@nobreak
    \everypar{}%
    \ifnum\LastMac=\Hae \addvspace{1pt plus 1pt minus .5pt}\fi
  \else
    \addpen{\gds@cbrk}%
    \addvspace{\onehalf}%
  \fi
  \bgroup
    \ninepoint\bf
    \Raggedright
    \global\advance\SecSec \@ne
    \noindent\@nohdbrk\thesubsection \hskip 1pc\relax #1\par
    \global\SecSecSec=\z@
  \egroup
  \nobreak
  \vskip\half
  \nobreak
  \@noafterindent
  \LastMac=\Hbe\relax
}

\def\@ssubsection#1{
  \if@nobreak
    \everypar{}%
    \ifnum\LastMac=\Hae \addvspace{1pt plus 1pt minus .5pt}\fi
  \else
    \addpen{\gds@cbrk}%
    \addvspace{\onehalf}%
  \fi
  \bgroup
    \ninepoint\bf
    \Raggedright
    \noindent\@nohdbrk #1\par
  \egroup
  \nobreak
  \vskip\half
  \nobreak
  \@noafterindent
  \LastMac=\Hbe\relax
}

\def\subsubsection{\@ifstar{\@ssubsubsection}{\@subsubsection}}

\def\@subsubsection#1{
  \if@nobreak
    \everypar{}%
    \ifnum\LastMac=\Hbe \addvspace{1pt plus 1pt minus .5pt}\fi
  \else
    \addpen{\gds@cbrk}%
    \addvspace{\onehalf}%
  \fi
  \bgroup
    \ninepoint\it
    \Raggedright
    \global\advance\SecSecSec \@ne
    \noindent\@nohdbrk\thesubsubsection \hskip 1pc\relax #1\par
  \egroup
  \nobreak
  \vskip\half
  \nobreak
  \@noafterindent
  \LastMac=\Hce\relax
}

\def\@ssubsubsection#1{
  \if@nobreak
    \everypar{}%
    \ifnum\LastMac=\Hbe \addvspace{1pt plus 1pt minus .5pt}\fi
  \else
    \addpen{\gds@cbrk}%
    \addvspace{\onehalf}%
  \fi
  \bgroup
    \ninepoint\it
    \Raggedright
    \noindent\@nohdbrk #1\par
  \egroup
  \nobreak
  \vskip\half
  \nobreak
  \@noafterindent
  \LastMac=\Hce\relax
}

\def\paragraph#1{
  \if@nobreak
    \everypar{}%
  \else
    \addpen{\gds@cbrk}%
    \addvspace{\one}%
  \fi%
  \bgroup%
    \ninepoint\it
    \noindent #1\ \nobreak%
  \egroup
  \LastMac=\Hde\relax
  \ignorespaces
}


\newif\ifappendix

\def\appendix{%
  \global\appendixtrue
  \def\thesection{\Alph{Sec}}%
  \def\thesubsection{\thesection\arabic{SecSec}}%
  \def\theeq{\thesection\arabic{Eqnno}}%
  \Sec=\z@ \SecSec=\z@ \SecSecSec=\z@ \Eqnno=\z@ \SubEqnno=\z@\relax
}




\def\beginlist{%
  \par\if@nobreak \else\addvspace{\half}\fi%
  \bgroup%
    \ninepoint
    \let\item=\list@item%
}

\def\list@item{%
  \par\noindent\hskip 1em\relax%
  \ignorespaces%
}

\def\endlist{\par\egroup\addvspace{\half}\@doendpe}


\def\beginrefs{%
  \par
  \bgroup
    \eightpoint
    \Fullout
    \let\bibitem=\bib@item
}

\def\bib@item{%
  \par\parindent=1.5em\Hang{1.5em}{1}%
  \everypar={\Hang{1.5em}{1}\ignorespaces}%
  \noindent\ignorespaces
}

\def\endrefs{\par\egroup\@doendpe}


\newtoks\CatchLine

\def\@journal{Mon.\ Not.\ R.\ Astron.\ Soc.\ }  
\def\@pubyear{1994}        
\def\@pagerange{000--000}  
\def\@volume{000}          
\def\@microfiche{}         %

\def\pubyear#1{\gdef\@pubyear{#1}\@makecatchline}
\def\pagerange#1{\gdef\@pagerange{#1}\@makecatchline}
\def\volume#1{\gdef\@volume{#1}\@makecatchline}
\def\microfiche#1{\gdef\@microfiche{and Microfiche\ #1}\@makecatchline}

\def\@makecatchline{%
  \global\CatchLine{%
    {\rm \@journal {\bf \@volume},\ \@pagerange\ (\@pubyear)\ \@microfiche}}%
}

\@makecatchline 

\newtoks\LeftHeader
\def\shortauthor#1{
  \global\LeftHeader{#1}%
}

\newtoks\RightHeader
\def\shorttitle#1{
  \global\RightHeader{#1}%
}

\def\PageHead{
  \begingroup
    \ifsp@page
      \csname ps@\sp@type\endcsname
    \fi
    \ifodd\pageno
      \let\the@head=\@oddhead
    \else
      \let\the@head=\@evenhead
    \fi
    \vbox to \z@{\vskip-22.5\p@%
      \hbox to \PageWidth{\vbox to8.5\p@{}%
        \the@head
      }%
    \vss}%
  \endgroup
  \nointerlineskip
}

\gdef\PageFoot{%
  \nointerlineskip%
  \begingroup
  \ifsp@page
    \csname ps@\sp@type\endcsname
    \global\sp@pagefalse
  \fi
  \vbox to 22pt{\vfil%
    \hbox to \PageWidth{%
      \eightpoint\strut\noindent
      \ifodd\pageno
        \@oddfoot
      \else
        \@evenfoot
      \fi
    }%
  }%
  \endgroup
}

\def\today{%
  \number\day\space
  \ifcase\month\or January\or February\or March\or April\or May\or June\or
    July\or August\or September\or October\or November\or December\fi
  \space\number\year%
}

\def\authorcomment#1{%
  \gdef\PageFoot{%
    \nointerlineskip%
    \vbox to 20pt{\vfil%
      \hbox to \PageWidth{\elevenpoint\noindent \hfil #1 \hfil}}%
  }%
}


\newif\ifplate@page
\newbox\plt@box

\def\beginplatepage{%
  \let\plate=\plate@head
  \let\caption=\fig@caption
  \global\setbox\plt@box=\vbox\bgroup
  \TEMPDIMEN=\PageWidth 
  \hsize=\PageWidth\relax
}

\def\endplatepage{\par\egroup\global\plate@pagetrue}
\def\plate@head#1{\gdef\plt@cap{#1}}


\def\letters{%
  \gdef\folio{\ifnum\pageno<\z@ L\romannumeral-\pageno
    \else L\number\pageno \fi}%
}


\newdimen\mathindent

\global\mathindent=\z@
\global\everydisplay{\global\@dspwd=\displaywidth\displaysetup}


\def\@displaylines#1{
  {}$\displ@y\hbox{\vbox{\halign{$\@lign\hfil\displaystyle##\hfil$\crcr
  #1\crcr}}}${}%
}

\def\@eqalign#1{\null\vcenter{\openup\jot\m@th
  \ialign{\strut\hfil$\displaystyle{##}$&$\displaystyle{{}##}$\hfil
      \crcr#1\crcr}}%
}

\def\@eqalignno#1{
  \global\advance\@dspwd by -\mathindent%
  {}$\displ@y\hbox{\vbox{\halign to\@dspwd%
  {\hfil$\@lign\displaystyle{##}$\tabskip\z@skip
  &$\@lign\displaystyle{{}##}$\hfil\tabskip\centering
  &\llap{$\@lign##$}\tabskip\z@skip\crcr
  #1\crcr}}}${}%
}


\global\let\displaylines=\@displaylines
\global\let\eqalign=\@eqalign
\global\let\eqalignno=\@eqalignno
\global\let\leqalignno=\@eqalignno

\newdimen\@dspwd   \@dspwd=\z@
\newif\if@eqno
\newif\if@leqno
\newtoks\@eqn
\newtoks\@eq

\def\displaysetup#1$${\displaytest#1\eqno\eqno\displaytest}

\def\displaytest#1\eqno#2\eqno#3\displaytest{%
 \if!#3!\ldisplaytest#1\leqno\leqno\ldisplaytest
 \else\@eqnotrue\@leqnofalse\@eqn={#2}\@eq={#1}\fi
 \generaldisplay$$}

\def\ldisplaytest#1\leqno#2\leqno#3\ldisplaytest{%
\@eq={#1}%
 \if!#3!\@eqnofalse\else\@eqnotrue\@leqnotrue
  \@eqn={#2}\fi}

\def\generaldisplay{%
  \if@eqno
    \if@leqno
      \hbox to \displaywidth{\noindent
        \rlap{$\displaystyle\the\@eqn$}%
        \hskip\mathindent$\displaystyle\the\@eq$\hfil}%
    \else
      \hbox to \displaywidth{\noindent
        \hskip\mathindent
        $\displaystyle\the\@eq$\hfil$\displaystyle\the\@eqn$}%
    \fi
  \else
    \hbox to \displaywidth{\noindent
      \hskip\mathindent$\displaystyle\the\@eq$\hfil}%
  \fi
}


\def\@notice{%
  \par\Two%
  \noindent{\b@ls{11pt}\ninerm This paper has been produced using the
    Royal Astronomical Society/Blackwell Science \TeX\ macros.\par}%
}

\outer\def\bye{\@notice\par\vfill\supereject\end}


\def\start@mess{%
  Monthly notices of the RAS journal style (\@typeface)\space
    v\@version,\space \@verdate.%
}

\everyjob{\Warn{\start@mess}}



\newif\if@debug \@debugfalse  

\def\Print#1{\if@debug\immediate\write16{#1}\else \fi}
\def\Warn#1{\immediate\write16{#1}}
\def\wlog#1{}

\newcount\Iteration 

\def\Single{0} \def\Double{1}                 
\def\Figure{0} \def\Table{1}                  

\def\InStack{0}  
\def\InZoneA{1}
\def\InZoneB{2}
\def\InZoneC{3}

\newcount\TEMPCOUNT 
\newdimen\TEMPDIMEN 
\newbox\TEMPBOX     
\newbox\VOIDBOX     

\newcount\LengthOfStack 
\newcount\MaxItems      
\newcount\StackPointer
\newcount\Point         
\newcount\NextFigure    
\newcount\NextTable     
\newcount\NextItem      

\newcount\StatusStack   
\newcount\NumStack      
\newcount\TypeStack     
\newcount\SpanStack     
\newcount\BoxStack      

\newcount\ItemSTATUS    
\newcount\ItemNUMBER    
\newcount\ItemTYPE      
\newcount\ItemSPAN      
\newbox\ItemBOX         
\newdimen\ItemSIZE      

\newdimen\PageHeight    
\newdimen\TextLeading   
\newdimen\Feathering    
\newcount\LinesPerPage  
\newdimen\ColumnWidth   
\newdimen\ColumnGap     
\newdimen\PageWidth     
\newdimen\BodgeHeight   
\newcount\Leading       

\newdimen\ZoneBSize  
\newdimen\TextSize   
\newbox\ZoneABOX     
\newbox\ZoneBBOX     
\newbox\ZoneCBOX     

\newif\ifFirstSingleItem
\newif\ifFirstZoneA
\newif\ifMakePageInComplete
\newif\ifMoreFigures \MoreFiguresfalse 
\newif\ifMoreTables  \MoreTablesfalse  

\newif\ifFigInZoneB 
\newif\ifFigInZoneC 
\newif\ifTabInZoneB 
\newif\ifTabInZoneC

\newif\ifZoneAFullPage

\newbox\MidBOX    
\newbox\LeftBOX
\newbox\RightBOX
\newbox\PageBOX   

\newif\ifLeftCOL  
\LeftCOLtrue

\newdimen\ZoneBAdjust

\newcount\ItemFits
\def\Yes{1}
\def\No{2}


\MaxItems=15
\NextFigure=\z@        
\NextTable=\@ne

\BodgeHeight=6pt
\TextLeading=11pt    
\Leading=11
\Feathering=\z@      
\LinesPerPage=61     
\topskip=\TextLeading
\ColumnWidth=20pc    
\ColumnGap=2pc       

\newskip\ItemSepamount  
\ItemSepamount=\TextLeading plus \TextLeading minus 4pt

\parskip=\z@ plus .1pt
\parindent=18pt
\widowpenalty=\z@
\clubpenalty=10000
\tolerance=1500
\hbadness=1500
\abovedisplayskip=6pt plus 2pt minus 1pt
\belowdisplayskip=6pt plus 2pt minus 1pt
\abovedisplayshortskip=6pt plus 2pt minus 1pt
\belowdisplayshortskip=6pt plus 2pt minus 1pt

\frenchspacing

\ninepoint 

\PageHeight=682pt
\PageWidth=2\ColumnWidth
\advance\PageWidth by \ColumnGap

\pagestyle{headings}




\newcount\DUMMY \StatusStack=\allocationnumber
\newcount\DUMMY \newcount\DUMMY \newcount\DUMMY 
\newcount\DUMMY \newcount\DUMMY \newcount\DUMMY 
\newcount\DUMMY \newcount\DUMMY \newcount\DUMMY
\newcount\DUMMY \newcount\DUMMY \newcount\DUMMY 
\newcount\DUMMY \newcount\DUMMY \newcount\DUMMY

\newcount\DUMMY \NumStack=\allocationnumber
\newcount\DUMMY \newcount\DUMMY \newcount\DUMMY 
\newcount\DUMMY \newcount\DUMMY \newcount\DUMMY 
\newcount\DUMMY \newcount\DUMMY \newcount\DUMMY 
\newcount\DUMMY \newcount\DUMMY \newcount\DUMMY 
\newcount\DUMMY \newcount\DUMMY \newcount\DUMMY

\newcount\DUMMY \TypeStack=\allocationnumber
\newcount\DUMMY \newcount\DUMMY \newcount\DUMMY 
\newcount\DUMMY \newcount\DUMMY \newcount\DUMMY 
\newcount\DUMMY \newcount\DUMMY \newcount\DUMMY 
\newcount\DUMMY \newcount\DUMMY \newcount\DUMMY 
\newcount\DUMMY \newcount\DUMMY \newcount\DUMMY

\newcount\DUMMY \SpanStack=\allocationnumber
\newcount\DUMMY \newcount\DUMMY \newcount\DUMMY 
\newcount\DUMMY \newcount\DUMMY \newcount\DUMMY 
\newcount\DUMMY \newcount\DUMMY \newcount\DUMMY 
\newcount\DUMMY \newcount\DUMMY \newcount\DUMMY 
\newcount\DUMMY \newcount\DUMMY \newcount\DUMMY

\newbox\DUMMY   \BoxStack=\allocationnumber
\newbox\DUMMY   \newbox\DUMMY \newbox\DUMMY 
\newbox\DUMMY   \newbox\DUMMY \newbox\DUMMY 
\newbox\DUMMY   \newbox\DUMMY \newbox\DUMMY 
\newbox\DUMMY   \newbox\DUMMY \newbox\DUMMY 
\newbox\DUMMY   \newbox\DUMMY \newbox\DUMMY

\def\wlog{\immediate\write\m@ne}


\def\GetItemAll#1{%
 \GetItemSTATUS{#1}
 \GetItemNUMBER{#1}
 \GetItemTYPE{#1}
 \GetItemSPAN{#1}
 \GetItemBOX{#1}
}

\def\GetItemSTATUS#1{%
 \Point=\StatusStack
 \advance\Point by #1
 \global\ItemSTATUS=\count\Point
}

\def\GetItemNUMBER#1{%
 \Point=\NumStack
 \advance\Point by #1
 \global\ItemNUMBER=\count\Point
}

\def\GetItemTYPE#1{%
 \Point=\TypeStack
 \advance\Point by #1
 \global\ItemTYPE=\count\Point
}

\def\GetItemSPAN#1{%
 \Point\SpanStack
 \advance\Point by #1
 \global\ItemSPAN=\count\Point
}

\def\GetItemBOX#1{%
 \Point=\BoxStack
 \advance\Point by #1
 \global\setbox\ItemBOX=\vbox{\copy\Point}
 \global\ItemSIZE=\ht\ItemBOX
 \global\advance\ItemSIZE by \dp\ItemBOX
 \TEMPCOUNT=\ItemSIZE
 \divide\TEMPCOUNT by \Leading
 \divide\TEMPCOUNT by 65536
 \advance\TEMPCOUNT \@ne
 \ItemSIZE=\TEMPCOUNT pt
 \global\multiply\ItemSIZE by \Leading
}


\def\JoinStack{%
 \ifnum\LengthOfStack=\MaxItems 
  \Warn{WARNING: Stack is full...some items will be lost!}
 \else
  \Point=\StatusStack
  \advance\Point by \LengthOfStack
  \global\count\Point=\ItemSTATUS
  \Point=\NumStack
  \advance\Point by \LengthOfStack
  \global\count\Point=\ItemNUMBER
  \Point=\TypeStack
  \advance\Point by \LengthOfStack
  \global\count\Point=\ItemTYPE
  \Point\SpanStack
  \advance\Point by \LengthOfStack
  \global\count\Point=\ItemSPAN
  \Point=\BoxStack
  \advance\Point by \LengthOfStack
  \global\setbox\Point=\vbox{\copy\ItemBOX}
  \global\advance\LengthOfStack \@ne
  \ifnum\ItemTYPE=\Figure 
   \global\MoreFigurestrue
  \else
   \global\MoreTablestrue
  \fi
 \fi
}


\def\LeaveStack#1{%
 {\Iteration=#1
 \loop
 \ifnum\Iteration<\LengthOfStack
  \advance\Iteration \@ne
  \GetItemSTATUS{\Iteration}
   \advance\Point by \m@ne
   \global\count\Point=\ItemSTATUS
  \GetItemNUMBER{\Iteration}
   \advance\Point by \m@ne
   \global\count\Point=\ItemNUMBER
  \GetItemTYPE{\Iteration}
   \advance\Point by \m@ne
   \global\count\Point=\ItemTYPE
  \GetItemSPAN{\Iteration}
   \advance\Point by \m@ne
   \global\count\Point=\ItemSPAN
  \GetItemBOX{\Iteration}
   \advance\Point by \m@ne
   \global\setbox\Point=\vbox{\copy\ItemBOX}
 \repeat}
 \global\advance\LengthOfStack by \m@ne
}


\newif\ifStackNotClean

\def\CleanStack{%
 \StackNotCleantrue
 {\Iteration=\z@
  \loop
   \ifStackNotClean
    \GetItemSTATUS{\Iteration}
    \ifnum\ItemSTATUS=\InStack
     \advance\Iteration \@ne
     \else
      \LeaveStack{\Iteration}
    \fi
   \ifnum\LengthOfStack<\Iteration
    \StackNotCleanfalse
   \fi
 \repeat}
}


\def\FindItem#1#2{%
 \global\StackPointer=\m@ne 
 {\Iteration=\z@
  \loop
  \ifnum\Iteration<\LengthOfStack
   \GetItemSTATUS{\Iteration}
   \ifnum\ItemSTATUS=\InStack
    \GetItemTYPE{\Iteration}
    \ifnum\ItemTYPE=#1
     \GetItemNUMBER{\Iteration}
     \ifnum\ItemNUMBER=#2
      \global\StackPointer=\Iteration
      \Iteration=\LengthOfStack 
     \fi
    \fi
   \fi
  \advance\Iteration \@ne
 \repeat}
}


\def\FindNext{%
 \global\StackPointer=\m@ne 
 {\Iteration=\z@
  \loop
  \ifnum\Iteration<\LengthOfStack
   \GetItemSTATUS{\Iteration}
   \ifnum\ItemSTATUS=\InStack
    \GetItemTYPE{\Iteration}
   \ifnum\ItemTYPE=\Figure
    \ifMoreFigures
      \global\NextItem=\Figure
      \global\StackPointer=\Iteration
      \Iteration=\LengthOfStack 
    \fi
   \fi
   \ifnum\ItemTYPE=\Table
    \ifMoreTables
      \global\NextItem=\Table
      \global\StackPointer=\Iteration
      \Iteration=\LengthOfStack 
    \fi
   \fi
  \fi
  \advance\Iteration \@ne
 \repeat}
}


\def\ChangeStatus#1#2{%
 \Point=\StatusStack
 \advance\Point by #1
 \global\count\Point=#2
}



\def\Zone{\InZoneA}

\ZoneBAdjust=\z@

\def\MakePage{
 \global\ZoneBSize=\PageHeight
 \global\TextSize=\ZoneBSize
 \global\ZoneAFullPagefalse
 \global\topskip=\TextLeading
 \MakePageInCompletetrue
 \MoreFigurestrue
 \MoreTablestrue
 \FigInZoneBfalse
 \FigInZoneCfalse
 \TabInZoneBfalse
 \TabInZoneCfalse
 \global\FirstSingleItemtrue
 \global\FirstZoneAtrue
 \global\setbox\ZoneABOX=\box\VOIDBOX
 \global\setbox\ZoneBBOX=\box\VOIDBOX
 \global\setbox\ZoneCBOX=\box\VOIDBOX
 \loop
  \ifMakePageInComplete
 \FindNext
 \ifnum\StackPointer=\m@ne
  \NextItem=\m@ne
  \MoreFiguresfalse
  \MoreTablesfalse
 \fi
 \ifnum\NextItem=\Figure
   \FindItem{\Figure}{\NextFigure}
   \ifnum\StackPointer=\m@ne \global\MoreFiguresfalse
   \else
    \GetItemSPAN{\StackPointer}
    \ifnum\ItemSPAN=\Single \def\Zone{\InZoneB}\relax
     \ifFigInZoneC \global\MoreFiguresfalse\fi
    \else
     \def\Zone{\InZoneA}
     \ifFigInZoneB \def\Zone{\InZoneC}\fi
    \fi
   \fi
   \ifMoreFigures\Print{}\FigureItems\fi
 \fi
\ifnum\NextItem=\Table
   \FindItem{\Table}{\NextTable}
   \ifnum\StackPointer=\m@ne \global\MoreTablesfalse
   \else
    \GetItemSPAN{\StackPointer}
    \ifnum\ItemSPAN=\Single\relax
     \ifTabInZoneC \global\MoreTablesfalse\fi
    \else
     \def\Zone{\InZoneA}
     \ifTabInZoneB \def\Zone{\InZoneC}\fi
    \fi
   \fi
   \ifMoreTables\Print{}\TableItems\fi
 \fi
   \MakePageInCompletefalse 
   \ifMoreFigures\MakePageInCompletetrue\fi
   \ifMoreTables\MakePageInCompletetrue\fi
 \repeat
 \ifZoneAFullPage
  \global\TextSize=\z@
  \global\ZoneBSize=\z@
  \global\vsize=\z@\relax
  \global\topskip=\z@\relax
  \vbox to \z@{\vss}
  \eject
 \else
 \global\advance\ZoneBSize by -\ZoneBAdjust
 \global\vsize=\ZoneBSize
 \global\hsize=\ColumnWidth
 \global\ZoneBAdjust=\z@
 \ifdim\TextSize<23pt
 \Warn{}
 \Warn{* Making column fall short: TextSize=\the\TextSize *}
 \vskip-\lastskip\eject\fi
 \fi
}

\def\MakeRightCol{
 \global\TextSize=\ZoneBSize
 \MakePageInCompletetrue
 \MoreFigurestrue
 \MoreTablestrue
 \global\FirstSingleItemtrue
 \global\setbox\ZoneBBOX=\box\VOIDBOX
 \def\Zone{\InZoneB}
 \loop
  \ifMakePageInComplete
 \FindNext
 \ifnum\StackPointer=\m@ne
  \NextItem=\m@ne
  \MoreFiguresfalse
  \MoreTablesfalse
 \fi
 \ifnum\NextItem=\Figure
   \FindItem{\Figure}{\NextFigure}
   \ifnum\StackPointer=\m@ne \MoreFiguresfalse
   \else
    \GetItemSPAN{\StackPointer}
    \ifnum\ItemSPAN=\Double\relax
     \MoreFiguresfalse\fi
   \fi
   \ifMoreFigures\Print{}\FigureItems\fi
 \fi
 \ifnum\NextItem=\Table
   \FindItem{\Table}{\NextTable}
   \ifnum\StackPointer=\m@ne \MoreTablesfalse
   \else
    \GetItemSPAN{\StackPointer}
    \ifnum\ItemSPAN=\Double\relax
     \MoreTablesfalse\fi
   \fi
   \ifMoreTables\Print{}\TableItems\fi
 \fi
   \MakePageInCompletefalse 
   \ifMoreFigures\MakePageInCompletetrue\fi
   \ifMoreTables\MakePageInCompletetrue\fi
 \repeat
 \ifZoneAFullPage
  \global\TextSize=\z@
  \global\ZoneBSize=\z@
  \global\vsize=\z@\relax
  \global\topskip=\z@\relax
  \vbox to \z@{\vss}
  \eject
 \else
 \global\vsize=\ZoneBSize
 \global\hsize=\ColumnWidth
 \ifdim\TextSize<23pt
 \Warn{}
 \Warn{* Making column fall short: TextSize=\the\TextSize *}
 \vskip-\lastskip\eject\fi
\fi
}

\def\FigureItems{
 \Print{Considering...}
 \ShowItem{\StackPointer}
 \GetItemBOX{\StackPointer} 
 \GetItemSPAN{\StackPointer}
  \CheckFitInZone 
  \ifnum\ItemFits=\Yes
   \ifnum\ItemSPAN=\Single
     \ChangeStatus{\StackPointer}{\InZoneB} 
     \global\FigInZoneBtrue
     \ifFirstSingleItem
      \hbox{}\vskip-\BodgeHeight
     \global\advance\ItemSIZE by \TextLeading
     \fi
     \unvbox\ItemBOX\ItemSep
     \global\FirstSingleItemfalse
     \global\advance\TextSize by -\ItemSIZE
     \global\advance\TextSize by -\TextLeading
   \else
    \ifFirstZoneA
     \global\advance\ItemSIZE by \TextLeading
     \global\FirstZoneAfalse\fi
    \global\advance\TextSize by -\ItemSIZE
    \global\advance\TextSize by -\TextLeading
    \global\advance\ZoneBSize by -\ItemSIZE
    \global\advance\ZoneBSize by -\TextLeading
    \ifFigInZoneB\relax
     \else
     \ifdim\TextSize<3\TextLeading
     \global\ZoneAFullPagetrue
     \fi
    \fi
    \ChangeStatus{\StackPointer}{\Zone}
    \ifnum\Zone=\InZoneC \global\FigInZoneCtrue\fi
  \fi
   \Print{TextSize=\the\TextSize}
   \Print{ZoneBSize=\the\ZoneBSize}
  \global\advance\NextFigure \@ne
   \Print{This figure has been placed.}
  \else
   \Print{No space available for this figure...holding over.}
   \Print{}
   \global\MoreFiguresfalse
  \fi
}

\def\TableItems{
 \Print{Considering...}
 \ShowItem{\StackPointer}
 \GetItemBOX{\StackPointer} 
 \GetItemSPAN{\StackPointer}
  \CheckFitInZone 
  \ifnum\ItemFits=\Yes
   \ifnum\ItemSPAN=\Single
    \ChangeStatus{\StackPointer}{\InZoneB}
     \global\TabInZoneBtrue
     \ifFirstSingleItem
      \hbox{}\vskip-\BodgeHeight
     \global\advance\ItemSIZE by \TextLeading
     \fi
     \unvbox\ItemBOX\ItemSep
     \global\FirstSingleItemfalse
     \global\advance\TextSize by -\ItemSIZE
     \global\advance\TextSize by -\TextLeading
   \else
    \ifFirstZoneA
    \global\advance\ItemSIZE by \TextLeading
    \global\FirstZoneAfalse\fi
    \global\advance\TextSize by -\ItemSIZE
    \global\advance\TextSize by -\TextLeading
    \global\advance\ZoneBSize by -\ItemSIZE
    \global\advance\ZoneBSize by -\TextLeading
    \ifFigInZoneB\relax
     \else
     \ifdim\TextSize<3\TextLeading
     \global\ZoneAFullPagetrue
     \fi
    \fi
    \ChangeStatus{\StackPointer}{\Zone}
    \ifnum\Zone=\InZoneC \global\TabInZoneCtrue\fi
   \fi
  \global\advance\NextTable \@ne
   \Print{This table has been placed.}
  \else
  \Print{No space available for this table...holding over.}
   \Print{}
   \global\MoreTablesfalse
  \fi
}


\def\CheckFitInZone{%
{\advance\TextSize by -\ItemSIZE
 \advance\TextSize by -\TextLeading
 \ifFirstSingleItem
  \advance\TextSize by \TextLeading
 \fi
 \ifnum\Zone=\InZoneA\relax
  \else \advance\TextSize by -\ZoneBAdjust
 \fi
 \ifdim\TextSize<3\TextLeading \global\ItemFits=\No
 \else \global\ItemFits=\Yes\fi}
}

\def\BeginOpening{%
  \ninepoint
  \thispagestyle{titlepage}%
  \global\setbox\ItemBOX=\vbox\bgroup%
    \hsize=\PageWidth%
    \hrule height \z@
    \ifsinglecol\vskip 6pt\fi 
}

\let\begintopmatter=\BeginOpening  

\def\EndOpening{%
  \One
  \egroup
  \ifsinglecol
    \box\ItemBOX%
    \vskip\TextLeading plus 2\TextLeading
    \@noafterindent
  \else
    \ItemNUMBER=\z@%
    \ItemTYPE=\Figure
    \ItemSPAN=\Double
    \ItemSTATUS=\InStack
    \JoinStack
  \fi
}


\newif\if@here  \@herefalse

\def\no@float{\global\@heretrue}
\let\nofloat=\relax 

\def\beginfigure{%
  \@ifstar{\global\@dfloattrue \@bfigure}{\global\@dfloatfalse \@bfigure}%
}

\def\@bfigure#1{%
  \par
  \if@dfloat
    \ItemSPAN=\Double
    \TEMPDIMEN=\PageWidth
  \else
    \ItemSPAN=\Single
    \TEMPDIMEN=\ColumnWidth
  \fi
  \ifsinglecol
    \TEMPDIMEN=\PageWidth
  \else
    \ItemSTATUS=\InStack
    \ItemNUMBER=#1%
    \ItemTYPE=\Figure
  \fi
  \bgroup
    \hsize=\TEMPDIMEN
    \global\setbox\ItemBOX=\vbox\bgroup
      \eightpoint\nostb@ls{10pt}%
      \let\caption=\fig@caption
      \ifsinglecol \let\nofloat=\no@float\fi
}

\def\fig@caption#1{%
  \vskip 5.5pt plus 6pt%
  \bgroup 
    \eightpoint\nostb@ls{10pt}%
    \setbox\TEMPBOX=\hbox{#1}%
    \ifdim\wd\TEMPBOX>\TEMPDIMEN
      \noindent \unhbox\TEMPBOX\par
    \else
      \hbox to \hsize{\hfil\unhbox\TEMPBOX\hfil}%
    \fi
  \egroup
}

\def\endfigure{%
  \par\egroup 
  \egroup
  \ifsinglecol
    \if@here \midinsert\global\@herefalse\else \topinsert\fi
      \unvbox\ItemBOX
    \endinsert
  \else
    \JoinStack
    \Print{Processing source for figure \the\ItemNUMBER}%
  \fi
}


\newbox\tab@cap@box
\def\tab@caption#1{\global\setbox\tab@cap@box=\hbox{#1\par}}

\newtoks\tab@txt@toks
\long\def\tab@txt#1{\global\tab@txt@toks={#1}\global\table@txttrue}

\newif\iftable@txt  \table@txtfalse
\newif\if@dfloat    \@dfloatfalse

\def\begintable{%
  \@ifstar{\global\@dfloattrue \@btable}{\global\@dfloatfalse \@btable}%
}

\def\@btable#1{%
  \par
  \if@dfloat
    \ItemSPAN=\Double
    \TEMPDIMEN=\PageWidth
  \else
    \ItemSPAN=\Single
    \TEMPDIMEN=\ColumnWidth
  \fi
  \ifsinglecol
    \TEMPDIMEN=\PageWidth
  \else
    \ItemSTATUS=\InStack
    \ItemNUMBER=#1%
    \ItemTYPE=\Table
  \fi
  \bgroup
    \eightpoint\nostb@ls{10pt}%
    \global\setbox\ItemBOX=\vbox\bgroup
      \let\caption=\tab@caption
      \let\tabletext=\tab@txt
      \ifsinglecol \let\nofloat=\no@float\fi
}

\def\endtable{%
  \par\egroup 
  \egroup
  \setbox\TEMPBOX=\hbox to \TEMPDIMEN{%
    \eightpoint\nostb@ls{10pt}%
    \hss
    \vbox{%
      \hsize=\wd\ItemBOX
      \ifvoid\tab@cap@box
      \else
        \noindent\unhbox\tab@cap@box
        \vskip 5.5pt plus 6pt%
      \fi
      \box\ItemBOX
      \iftable@txt
        \vskip 10pt%
        \noindent\the\tab@txt@toks
        \global\table@txtfalse
      \fi
    }%
    \hss
  }%
  \ifsinglecol
    \if@here \midinsert\global\@herefalse\else \topinsert\fi
      \box\TEMPBOX
    \endinsert
  \else
    \global\setbox\ItemBOX=\box\TEMPBOX
    \JoinStack
    \Print{Processing source for table \the\ItemNUMBER}%
  \fi
}

\def\UnloadZoneA{%
\FirstZoneAtrue
 \Iteration=\z@
  \loop
   \ifnum\Iteration<\LengthOfStack
    \GetItemSTATUS{\Iteration}
    \ifnum\ItemSTATUS=\InZoneA
     \GetItemBOX{\Iteration}
     \ifFirstZoneA \vbox to \BodgeHeight{\vfil}%
     \FirstZoneAfalse\fi
     \unvbox\ItemBOX\ItemSep
     \LeaveStack{\Iteration}
     \else
     \advance\Iteration \@ne
   \fi
 \repeat
}

\def\UnloadZoneC{%
\Iteration=\z@
  \loop
   \ifnum\Iteration<\LengthOfStack
    \GetItemSTATUS{\Iteration}
    \ifnum\ItemSTATUS=\InZoneC
     \GetItemBOX{\Iteration}
     \ItemSep\unvbox\ItemBOX
     \LeaveStack{\Iteration}
     \else
     \advance\Iteration \@ne
   \fi
 \repeat
}


\def\ShowItem#1{
  {\GetItemAll{#1}
  \Print{\the#1:
  {TYPE=\ifnum\ItemTYPE=\Figure Figure\else Table\fi}
  {NUMBER=\the\ItemNUMBER}
  {SPAN=\ifnum\ItemSPAN=\Single Single\else Double\fi}
  {SIZE=\the\ItemSIZE}}}
}

\def\ShowStack{%
 \Print{}
 \Print{LengthOfStack = \the\LengthOfStack}
 \ifnum\LengthOfStack=\z@ \Print{Stack is empty}\fi
 \Iteration=\z@
 \loop
 \ifnum\Iteration<\LengthOfStack
  \ShowItem{\Iteration}
  \advance\Iteration \@ne
 \repeat
}

\def\B#1#2{%
\hbox{\vrule\kern-0.4pt\vbox to #2{%
\hrule width #1\vfill\hrule}\kern-0.4pt\vrule}
}


\newif\ifsinglecol   \singlecolfalse

\def\onecolumn{%
  \global\output={\singlecoloutput}%
  \global\hsize=\PageWidth
  \global\vsize=\PageHeight
  \global\ColumnWidth=\hsize
  \global\TextLeading=12pt
  \global\Leading=12
  \global\singlecoltrue
  \global\let\onecolumn=\relax
  \global\let\footnote=\sing@footnote
  \global\let\vfootnote=\sing@vfootnote
  \ninepoint 
  \message{(Single column)}%
}

\def\singlecoloutput{%
  \shipout\vbox{\PageHead\vbox to \PageHeight{\pagebody\vss}\PageFoot}%
  \advancepageno
  \ifplate@page
    \shipout\vbox{%
      \sp@pagetrue
      \def\sp@type{plate}%
      \global\plate@pagefalse
      \PageHead\vbox to \PageHeight{\unvbox\plt@box\vfil}\PageFoot%
    }%
    \message{[plate]}%
    \advancepageno
  \fi
  \ifnum\outputpenalty>-\@MM \else\dosupereject\fi%
}

\def\ItemSep{\vskip\ItemSepamount\relax}

\def\ItemSepbreak{\par\ifdim\lastskip<\ItemSepamount
  \removelastskip\penalty-200\ItemSep\fi%
}


\let\@@endinsert=\endinsert 

\def\endinsert{\egroup 
  \if@mid \dimen@\ht\z@ \advance\dimen@\dp\z@ \advance\dimen@12\p@
    \advance\dimen@\pagetotal \advance\dimen@-\pageshrink
    \ifdim\dimen@>\pagegoal\@midfalse\p@gefalse\fi\fi
  \if@mid \ItemSep\box\z@\ItemSepbreak
  \else\insert\topins{\penalty100 
    \splittopskip\z@skip
    \splitmaxdepth\maxdimen \floatingpenalty\z@
    \ifp@ge \dimen@\dp\z@
    \vbox to\vsize{\unvbox\z@\kern-\dimen@}
    \else \box\z@\nobreak\ItemSep\fi}\fi\endgroup%
}


\def\gobbleone#1{}
\def\gobbletwo#1#2{}
\let\footnote=\gobbletwo 
\let\vfootnote=\gobbleone

\def\sing@footnote#1{\let\@sf\empty 
  \ifhmode\edef\@sf{\spacefactor\the\spacefactor}\/\fi
  \hbox{$^{\hbox{\eightpoint #1}}$}\@sf\sing@vfootnote{#1}%
}

\def\sing@vfootnote#1{\insert\footins\bgroup\eightpoint\b@ls{9pt}%
  \interlinepenalty\interfootnotelinepenalty
  \splittopskip\ht\strutbox 
  \splitmaxdepth\dp\strutbox \floatingpenalty\@MM
  \leftskip\z@skip \rightskip\z@skip \spaceskip\z@skip \xspaceskip\z@skip
  \noindent $^{\scriptstyle\hbox{#1}}$\hskip 4pt%
    \footstrut\futurelet\next\fo@t%
}

\def\footnoterule{\kern-3\p@ \hrule height \z@ \kern 3\p@}

\skip\footins=19.5pt plus 12pt minus 1pt
\count\footins=1000
\dimen\footins=\maxdimen

\def\note#1#2{%
  \let\@sf=\empty \ifhmode\edef\@sf{\spacefactor\the\spacefactor}\/\fi
  #1\insert\footins\bgroup
    \eightpoint\b@ls{10pt}\rm
    \interlinepenalty\interfootnotelinepenalty
    \splitmaxdepth\dp\strutbox \floatingpenalty\@MM
    \leftskip\z@skip \rightskip\z@skip \spaceskip\z@skip \xspaceskip\z@skip
    \noindent\footstrut #1$\,$#2\strut\par
  \egroup
  \@sf\relax}


\def\landscape{%
  \global\TEMPDIMEN=\PageWidth
  \global\PageWidth=\PageHeight
  \global\PageHeight=\TEMPDIMEN
  \global\let\landscape=\relax
  \onecolumn
  \message{(landscape)}%
  \raggedbottom
}


\output{%
  \ifLeftCOL
    \global\setbox\LeftBOX=\vbox to \ZoneBSize{\box255\unvbox\ZoneBBOX
      \ifvoid\footins\else
        \vskip\skip\footins\unvbox\footins\fi
    }%
    \global\LeftCOLfalse
    \MakeRightCol
  \else
    \setbox\RightBOX=\vbox to \ZoneBSize{\box255\unvbox\ZoneBBOX
      \ifvoid\footins\else
        \vskip\skip\footins\unvbox\footins\fi
    }%
    \setbox\MidBOX=\hbox{\box\LeftBOX\hskip\ColumnGap\box\RightBOX}%
    \setbox\PageBOX=\vbox to \PageHeight{%
      \UnloadZoneA\box\MidBOX\UnloadZoneC}%
    \shipout\vbox{\PageHead\vbox to \PageHeight{\box\PageBOX\vss}\PageFoot}%
    \advancepageno
    \ifplate@page
      \shipout\vbox{%
        \sp@pagetrue
        \def\sp@type{plate}%
        \global\plate@pagefalse
        \PageHead\vbox to \PageHeight{\unvbox\plt@box\vfil}\PageFoot%
      }%
      \message{[plate]}%
      \advancepageno
    \fi
    \global\LeftCOLtrue
    \CleanStack
    \MakePage
  \fi
}


\Warn{\start@mess}

\newif\ifCUPmtplainloaded 
\ifprod@font
  \global\CUPmtplainloadedtrue
\fi

\def\mnmacrosloaded{} 

\catcode `\@=12 



\fi
\input psfig.sty

%

\newif\ifAMStwofonts

\ifCUPmtplainloaded \else
  \NewTextAlphabet{textbfit} {cmbxti10} {}
  \NewTextAlphabet{textbfss} {cmssbx10} {}
  \NewMathAlphabet{mathbfit} {cmbxti10} {} 
  \NewMathAlphabet{mathbfss} {cmssbx10} {} 
  \ifAMStwofonts
    \NewSymbolFont{upmath} {eurm10}
    \NewSymbolFont{AMSa} {msam10}
    \NewMathSymbol{\upi}     {0}{upmath}{19}
    \NewMathSymbol{\umu}     {0}{upmath}{16}
    \NewMathSymbol{\upartial}{0}{upmath}{40}
    \NewMathSymbol{\leqslant}{3}{AMSa}{36}
    \NewMathSymbol{\geqslant}{3}{AMSa}{3E}

    \let\leq=\leqslant \let\le=\leqslant
    \let\geq=\geqslant 
  \else
    \def\umu{\mu}
    \def\upi{\pi}
    \def\upartial{\partial}
  \fi
\fi


\pageoffset{-2.5pc}{0pc}

\loadboldmathnames
 



\def\2248{RE~J2248-511}
\def\1034{RE~J1034+396}
\def\H{1H0419-577}
\def\E1346{E1346+266}
\def\0437{RX~J0437.4-4711}

\def\ASCA{{\it ASCA\/}}
\def\SAX{{\it Beppo-SAX\/}}
\def\ROSAT{{\it ROSAT\/}}
\def\EXOSAT{{\it EXOSAT\/}}

\def\puch{Puchnarewicz}

\def\approx{\sim}
\def\x{X-ray}
\def\eg{{e.g.~}}
\def\ie{{i.e.~}}
\def\etal{et~al.~}
\def\eV{~e\kern-.15em V}                 
\def\keV{~ke\kern-.15em V}                
\def\nH{$N_{\rm H}$}
\def\nHint{$N_{\rm Hint}$}
\def\nHgal{$N_{\rm HGal}$}

\def\Ha{H$\alpha$}
\def\Hb{H$\beta$}
\def\ao{$\alpha_{\rm  opt}$}
\def\ax{$\alpha_{\rm  x}$}
\def\aox{$\alpha_{\rm ox}$}

\def\kms{~km~s$^{-1}$}
\def\cm2{~cm$^{-2}$}
\def\exp{\times 10^}
\def\cs{~count~s$^{-1}$}  	
\def\cks{~count~ks$^{-1}$}     
\def\ks{~ks}
\def\persec{~s$^{-1}$}		

\def\deg{$^\circ$}

\def\X2{$\chi^2$}
\def\Xv2{$\chi^2_\nu$}

\begintopmatter  

\title{\2248 - Not all variable, ultrasoft, X-ray AGN have narrow Balmer lines}
\author{A.~A.~Breeveld,$^1$ 
	E.~M.~Puchnarewicz$^1$ and
	C.~Otani$^2$} 
\affiliation{$^1$ Mullard Space Science Laboratory, University College London,
Holmbury St. Mary, Dorking, Surrey, RH5 6NT}
\affiliation{$^2$ The Institute of Physical and Chemical Research (RIKEN), 2-1
Hirosawa, Wako, Saitama 351-01, Japan}

\shortauthor{A.~A.~Breeveld \etal}
\shorttitle{\2248 - a variable, broad line, ultrasoft AGN}


\abstract   {We present \ASCA\ data on \2248, extending existing optical and
soft X-ray coverage to 10\keV, and monitoring the soft component. These data
show that, despite a very strong ultrasoft \x\ excess below 0.3\keV\ and  a
soft 0.3--2\keV\ spectral index in earlier \ROSAT\ data, the hard X-ray
spectrum ($\alpha\sim -0.8$; 0.6-10\keV) is typical of type 1 AGN, and the
soft component has since disappeared. Optical data taken at two different
epochs show that the big blue bump is also highly variable. The strength of the
ultrasoft X-ray component and the extreme variability in \2248\ are reminiscent
of the behaviour observed in many narrow line Seyfert 1s (NLS1s). However, the
high energy end of the \ROSAT\ spectrum, the \ASCA\ spectrum and the Balmer
line full widths at half maximum of $\sim$3000 km s$^{-1}$ in \2248\, are
typical of normal Seyfert 1 AGN.

The change in the soft X-ray spectrum as observed in the \ROSAT\ and \ASCA\
data is consistent with the behaviour of Galactic Black Hole Candidates (GBHCs)
as they move from a high to a low state, \ie a fall in the ultrasoft component
and a hardening of the X-ray continuum. This GBHC analogy has also been
proposed for NLS1s. Alternatively, the variability may be caused by opacity
changes in a hot, optically-thin corona which surrounds a cold, dense accretion
disc; this was first suggested by Guainazzi \etal for \H, an object which
shows remarkably similar properties to \2248. }

\keywords {Galaxies: Seyfert -- Galaxies: active -- Galaxies:
individual: \2248}

\maketitle  

\section{Introduction}

The soft-to-medium X-ray spectrum ($\sim$0.1--10\keV) of active galactic nuclei
(AGN) is dominated by two continuum components, the soft X-ray excess (\eg
Arnaud \etal 1985; Turner \& Pounds 1989; Walter \& Fink 1993) and a hard X-ray
power law (\eg Turner \& Pounds 1989; Comastri \etal 1992; Nandra \etal 1997).
The soft excess dominates below $\sim$0.5\keV\ and  is believed to be the high
energy tail of one `big bump' component, of which the optical/UV big blue bump
(BBB) is the low energy end (Walter \& Fink 1993). A relationship between the
X-ray spectrum of AGN and the velocity of the broad line regions (BLRs) has now
been well established. In quasars there is an anticorrelation  between the
width of the \Hb\ line and the slope of the soft X-ray spectrum in the \ROSAT\
band, in the sense that the narrowest lines are seen where the slopes are
softest (Laor \etal 1997, hereafter L97; \puch\ \etal 1992). Boller, Brandt \&
Fink (1996) also demonstrated that nature usually discriminates against high
velocity BLRs in ultrasoft AGN in the general AGN population. There is further
evidence that the Balmer lines also `respond' to the slope of the hard (\ASCA\
$\sim$2--10\keV) spectrum, again with the narrow-line AGN having softer hard
X-ray continua (Brandt, Mathur \& Elvis 1997).

The reason for the dependence of BLR velocity on the X-ray spectrum is not
known. It has been suggested that orientation effects or black hole mass may be
important (\eg Puchnarewicz \etal 1992; L97; Grupe \etal 1998, hereafter G98;
Nandra \etal 1997), although direct effects, \eg changes in the shape of
ionizing continuum affecting the structure of the BLR, may also be the cause
(Rees, Netzer \& Ferland 1989, Brandt \etal 1994, Puchnarewicz \etal 1997).
Possible effects caused by the relative strength of the big bump component,
which carries most of the ionizing continuum flux, are also poorly understood.
L97 found that the 2\keV\ flux is relatively weak compared to the optical
when the soft component is strong, which may be an important factor.

In this paper, we investigate the relative strength of the big bump and the
relationship between the X-ray spectrum and BLR velocity using optical and
X-ray data, including recently obtained ASCA data, for the Seyfert 1 galaxy
\2248\ (z = 0.101). This is a rare EUV-selected AGN which was first detected in
the \ROSAT\ Wide Field Camera (WFC) All Sky Survey (Pounds \etal 1993).  In the
associated  identification programme it was shown to have broad permitted lines
[H$\beta$ FWHM 2900\kms; equivalent width (EW) 50~\AA] (Mason \etal 1995;
hereafter M95) and a very blue optical spectrum ($\alpha_{\rm  opt} \sim 0.8$;
$\alpha$ is defined throughout as in $f_\nu \propto \nu^{\alpha}$, note there
is no implicit negative). In a later pointed PSPC observation the  0.1--2\keV\
spectrum overall was soft and steepened significantly at low energies ($E \la
0.3$\keV) to $\alpha =-3$ (\puch\ \etal 1995a; hereafter P95). Between 0.3 and
2.0\keV, the spectrum  was best fitted by $\alpha =-1.6$, which is consistent
with  the slope of a single power law fitted to the PSPC (0.1--2\keV) spectra
of other UV-bright samples (\eg Walter \& Fink 1993; Schartel \etal 1996). The
very strong ultrasoft (\ie $E < 0.3$\keV) component in \2248\ has shown signs
of significant variability (P95). 

Thus \2248\ has exhibited an unusual two-component soft X-ray spectrum. There
is a  significant steepening below 0.3\keV\ (hereafter referred to as the
`ultrasoft' component) and a comparatively normal spectral slope in the higher
energy part of the PSPC band (hereafter the `soft X-ray' component). It may
well be that the soft X-ray component in \2248\ is the same as the single
component observed by L97 in comparable AGN, with an independent ultrasoft
component superposed on top. To our knowledge, a strong ultrasoft component in
addition to a normal soft X-ray spectrum, has been seen in only two other type
1 AGN (1H0419-577, Guainazzi \etal 1998; and RXJ0437.4-4711, Wang \etal 1997).

\section{\textbfit{ASCA\/} observation, data analysis and results}

\2248\ was observed with \ASCA\ (Tanaka, Inoue \& Holt 1994) on 1997 May 17
using the two Solid-state Imaging Spectrometer CCD detectors (SIS0 and SIS1)
and the two Gas Imaging Spectrometer scintillation proportional counters (GIS2
and GIS3). The SIS detectors were operated in 1 CCD mode, using the best
calibrated chips (chip 1 for SIS0 and 3 for SIS1). Faint mode was used for the
entire observation, the data being converted to Bright2 mode before reduction
and analysis (this allows corrections for echo effect and dark frame error).
The GIS were operated in pulse height mode, and standard telemetry bit
assignments were used. No lower level discriminator was used.

The data were initially processed by the Goddard Space Flight Center Guest
Observer Facility using the Revision 2 screening criteria, resulting in
15.5\ks\ per SIS and 13\ks\ per GIS of acceptable data. We have used {\sevenrm
FTOOLS}, {\sevenrm XSELECT}  and {\sevenrm XSPEC} for the analysis and fitting,
and have followed the recommendations of the \ASCA\ Data Reduction Guide
Version 2.0.

\subsection{Position}

\2248\ is clearly detected in all four detectors. The position of the centroid
of the source in the SIS detectors is ${\rm RA}(2000)=22^{\rm h}~48^{\rm
m}~40\fs 7$, ${\rm  Dec.}(2000)=-51^\circ~09'~56\arcsec$ and in the GIS
detectors is ${\rm RA}(2000)=22^{\rm h}~48^{\rm m}~38^{\rm s}$, ${\rm 
Dec.}(2000)=-51^\circ~09'~20\arcsec$. These measurements compare well with the 
previous optical measurement of ${\rm RA}(2000)=22^{\rm h}~48^{\rm m}~41\fs 2$,
${\rm Dec.}(2000)=-51^\circ~09'~54\arcsec$ (M95), given that the error on the
GIS position is at least 50 arcsec. There is no  evidence that the source is
extended and there are no other significant \x\ sources in the field of view.

\beginfigure*{1}
\psfig{figure=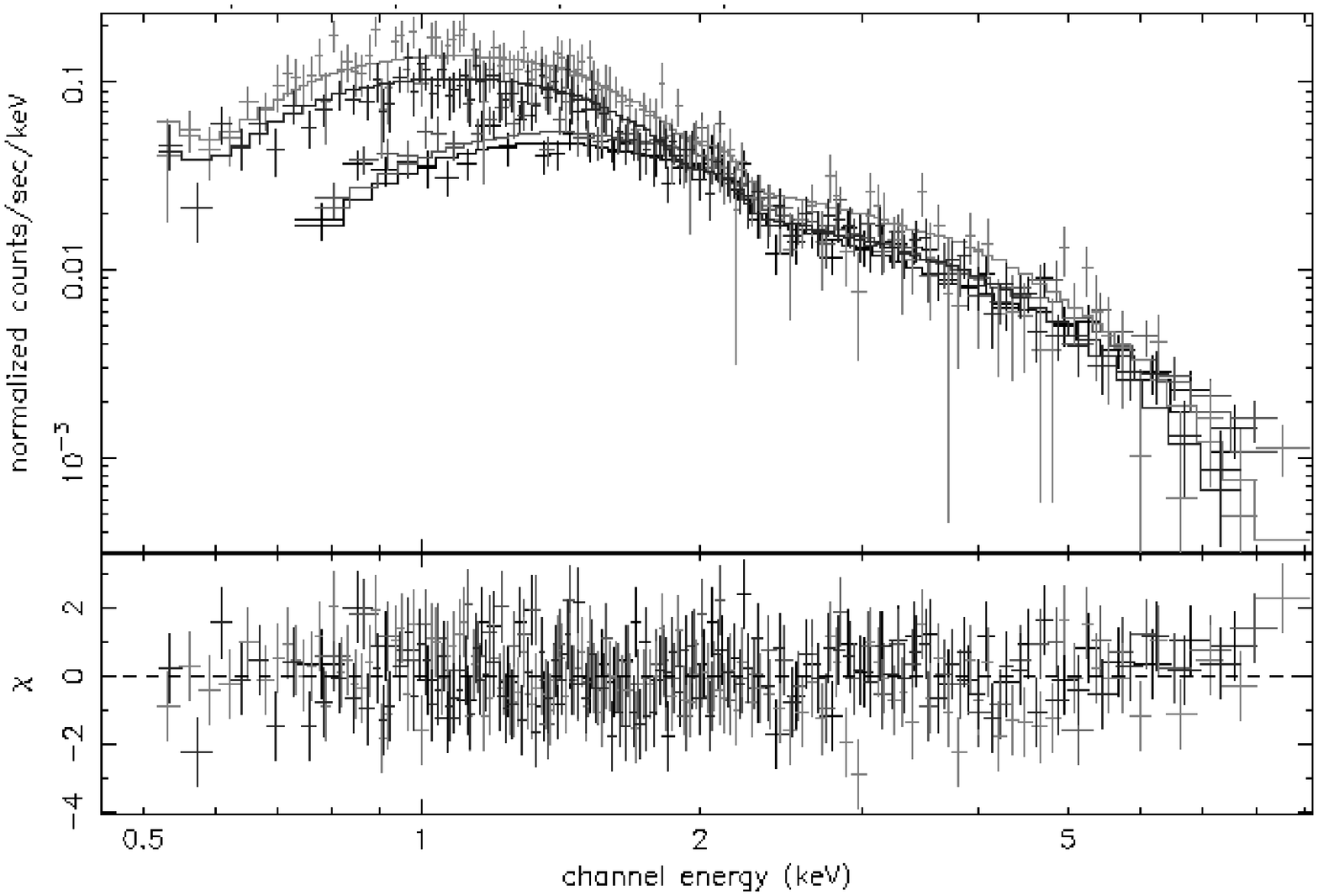,height=3.5in,width=6.5in,angle=270}
\caption{{\bf Figure 1.} The best fitting single power law fitted
simultaneously to the SIS and GIS data. The residuals to the fit, in the lower
window, are in units of $\Delta\chi^2$}
\endfigure

\subsection{Spectral analysis}

We used circular extraction regions, centred on the object, with radii of 4, 3,
6 and 6 arcmin for SIS0, SIS1, GIS2 and GIS3 respectively. A radius of 4 arcmin
is recommended for the SIS, but in the SIS1 the object lies too close to the
edge of the CCD to use a full 4 arcmin radius. Using a square region, or a
larger, but offset, circular region, instead of the 3 arcmin circle chosen,
does not affect the result significantly. We have carefully chosen circular,
source-free extraction regions for the background. For the SIS the background
is taken from the same chip as the source, whereas for the GIS, the background
is taken from a region at the same radius as the source from the optical axis
of the telescope.

We have extracted spectra from inside the specified extraction regions using
all of the acceptable exposure times. The SIS spectra were grouped to give 20
counts in each bin, and the GIS spectra 40 counts per bin. Suitable response
files were generated for each detector. Only events with grades 0,2,3,4 were
selected from the SIS data.

Using {\sevenrm XSPEC}, we first fitted the spectra from the two SIS detectors
simultaneously, and then from the two GIS detectors. The fits are given in
Table 1. The two sets of results were consistent within $1\sigma $, with the
GIS giving slightly harder slopes than the SIS. The data from all four
detectors were therefore simultaneously fitted to the same model, with the
relative normalizations allowed to go free, to allow for small calibration
uncertainties in the absolute normalization in each detector. We fixed the
local Galactic column at \nHgal$= 1.4 \exp{20}$\cm2 (interpolated from the maps
of Stark \etal 1984, see P95), and left the column of the host galaxy (the 
intrinsic column \nHint), at redshift 0.1, to be a free parameter, but greater
than or  equal to zero. A single, absorbed power law gave a good fit
(\Xv2$=0.89$) over the range 0.5--10.0\keV\ (SIS) 0.7--10.0\keV\ (GIS), with
$\alpha = -0.85^{+0.08}_{-0.05}$ and \nHint$ = 1.76\pm 4.4 \exp{20} $\cm2. The
data and fit are shown in Fig. 1. The errors $\Delta\alpha$ and $\Delta$\nHint\
in the fit were calculated by making a grid search for models with
$\Delta\chi^2=4.61$, as appropriate for a 90 per cent confidence level for two
interesting parameters (e.g. Press \etal 1989). Ignoring data below 1.0\keV\
leads to a slightly harder $\alpha = -0.83\pm 0.08$ although it is obviously
very close to the 0.5--10\keV\ fits.

An additional error comes from \nHgal\ itself. P95 use a 10 per cent
uncertainty, as calculated by Laor \etal 1994. Stark \etal (1992) suggest an
error 3 times larger than this, making our value for \nHgal$= 1.4 \exp{20}\pm 6
\exp{19}$\cm2. For safety, and because the declination is below $-40$\deg, we
have assumed the larger error and explored the effect on the fits given in
Table 1. We also found the 90 per cent confidence limits on the fit after
propagating through the maximum and minimum values for \nHgal. These limits
have been included in the table as a second set of errors where they are
outside the previous confidence limits. In the fitting to these \ASCA\ data,
the change in \nHgal\ has very little effect on $\alpha$.

In order to increase the spectral resolution at low energies to check for any
interesting features (\eg warm absorber edges), we combined the data from SIS0
and SIS1 in the manner described in the \ASCA\ guide to make a single SIS
spectrum. The background files and exposures were calculated carefully and the
final {\sc pha} file was grouped to 20 counts per bin. Again a single, absorbed
power law gave a good fit (\Xv2$=0.48$) with $\alpha = -0.83\pm 0.1$ and
\nHint$=0^{+4.6}_{-9.1}\exp{20}$\cm2, which is not significantly different from
the fit to all four detectors simultanously, as expected. No additional
features were required to fit the data. The reduced chi-squared values are very
low, implying that the error bars may be larger than necessary in the combined
data. This is probably associated with the method of combining the response
matrices. Never-the-less, in all the power-law fits so far described, the
slopes are consistent within their errors which suggests that the result is
robust (see Table 1). For convenience, the combined spectrum has been used
later on for comparison with the \ROSAT\ data.

George \etal (1998) comment that there are uncertainties in the XRT/SIS
effective area at energies below 0.6\keV\ and especially around 0.5\keV. For
this reason we have also included fits ignoring data below 0.6\keV\ to check
the consistency of the data (see Table 1). The \ASCA\ Guest Observer Facility
quantify this uncertainty and give a guideline of $2 \exp{20}$\cm2 as a 90
percent systematic error on the SIS for measurement of the line of sight column
density (see also Dotani \etal 1996). This factor has been taken into account
in the comparison between \ASCA\ and PSPC data (see Table 2 and Section
2.3.1).

We tried two component fits to the data to see if there were any break in the
power-law slope, or an excess at low energies. A broken power law or two power
laws gave no improvement in \X2. Adding a black body at 220\eV\ gave a small
improvement in \X2, but according to the F-test (see \eg Bevington 1969), the
improvement is not significant. 

\begintable*{1} 
   \caption{{\bf Table 1.} Power-law fits to ASCA GIS and SIS data}
   \halign{#\hfil\quad & #\hfil\quad &
   	\hfil#\hfil\quad & 
   	\hfil#\hfil\quad & 
   	\hfil#\hfil\quad & 
   	\hfil#\hfil\quad & 
   	\hfil#\hfil\cr
   		& Range (keV)	& \nHint$^a$ 		& $\alpha$ 	
		& Norm$^b$  	& \X2 /dof 		& \Xv2 \cr 
   SIS0+SIS1  	& 0.5--10.0 	& $2.1{^{+4.6}_{-4.2}}$ ${^{+5.3}_{-4.8}}$ 	
		& $-0.9^{+0.1}_{-0.09}$
 		& $1.4\pm0.1$ 	& $170/228$		& 0.74 \cr 
   GIS2+GIS3  	& 0.7--10.0 	& $0^{+0}_{-25}$ 	& $-0.87\pm 0.09$ 	
		& $1.5\pm0.1$ 	& $111/113$		& 0.98 \cr 
   SIS+GIS  	& 0.5(SIS)/0.7(GIS)--10.0 	
		& $1.76^{+3.7}_{-3.5}$ ${^{+4.4}_{-4.1}}$ 
						& $-0.85^{+0.08}_{-0.05}$ 	
		& $1.4\pm0.1$ 	& $307/346$		& 0.89 \cr 
   SIS+GIS  	& 1.0--10.0 	& $0^{+0}_{-22}$ 	& $-0.83\pm 0.08$ 	
		& $1.3\pm0.1$ 	& $246/288$		& 0.85 \cr 
   SIS combined	& 0.5--10.0	& $0{^{+3.5}_{-4}}$ ${^{+4.1}_{-4.8}}$
		& $-0.83 \pm 0.12$
		& $1.5 \pm 0.05$	& $86/178$		& 0.48  \cr
   SIS combined	& 0.6--10.0	& $0{^{+4.3}_{-8.6}}$ ${^{+4.6}_{-9.1}}$
		& $-0.83\pm 0.1$
		& $1.5 \pm 0.05$	& $82/171$		& 0.48  \cr
   SIS combined	& 1.0--10.0	& $0^{+0}_{-40}$	& $-0.83 \pm 0.12$
		& $1.5 \pm 0.1$	& $65/144$		& 0.45  \cr
    } 
   \tabletext{A value of $1.4 \exp{20}\pm 6 \exp{19}$\cm2 has been assumed for
the Galactic column density.

\noindent The errors were calculated by making a grid search at the 90 per cent
confidence level (\ie with $\Delta\chi^2=4.61$), for two interesting
parameters. The second set of errors, where given, are calculated assuming the
maximum and minimum values for \nHgal.

\noindent $^a \exp{20}$\cm2. The value is constrained to be greater than zero
although the errors have been allowed to go below zero.
	
\noindent $^b \exp{-3}$ photons \cm2\persec\ at 1\keV

}
\endtable

\subsubsection{FeK$\alpha$ line}

In almost all Seyfert 1 AGN observed with \ASCA\ an FeK$\alpha$
fluorescent line is significantly detected at 6.4\keV\ (neutral) or 6.7\keV\
(ionized), with an average EW of $121\pm 16$\eV\ (Nandra \etal 1997).  The EW
tends to be smaller for higher luminosity objects and these lines are rarely
seen in quasars. 

To test for FeK$\alpha$ emission in \2248, we added an unresolved Gaussian
emission line to the power-law fit to the \ASCA\ simultaneous SIS and GIS data.
According to the F-test, the additional component did
not provide a significant improvement at either 6.4 or 6.7\keV. Allowing the
width of the Gaussian line to vary as a free parameter gave an unrealistically
large FWHM and EW (13\keV\ FWHM and 6\keV\ EW) while making the underlying
power-law continuum very steep.  The upper limits (90 per cent) on the EW of a
narrow line added at 6.4 or 6.7\keV\ are 220\eV\ and 335\eV, respectively. 
These are above the average given by Nandra \etal (1997), although a lack of
significant FeK$\alpha$ emission would not be unexpected since the luminosity
of \2248\ ($L\sim10^{45}$ erg s$^{-1}$; $H_0$=50 km s$^{-1}$ Mpc$^{-1}$,
$q_0$=0) places it on the borderline between a Seyfert 1 and quasar.


\beginfigure*{2}
\psfig{figure=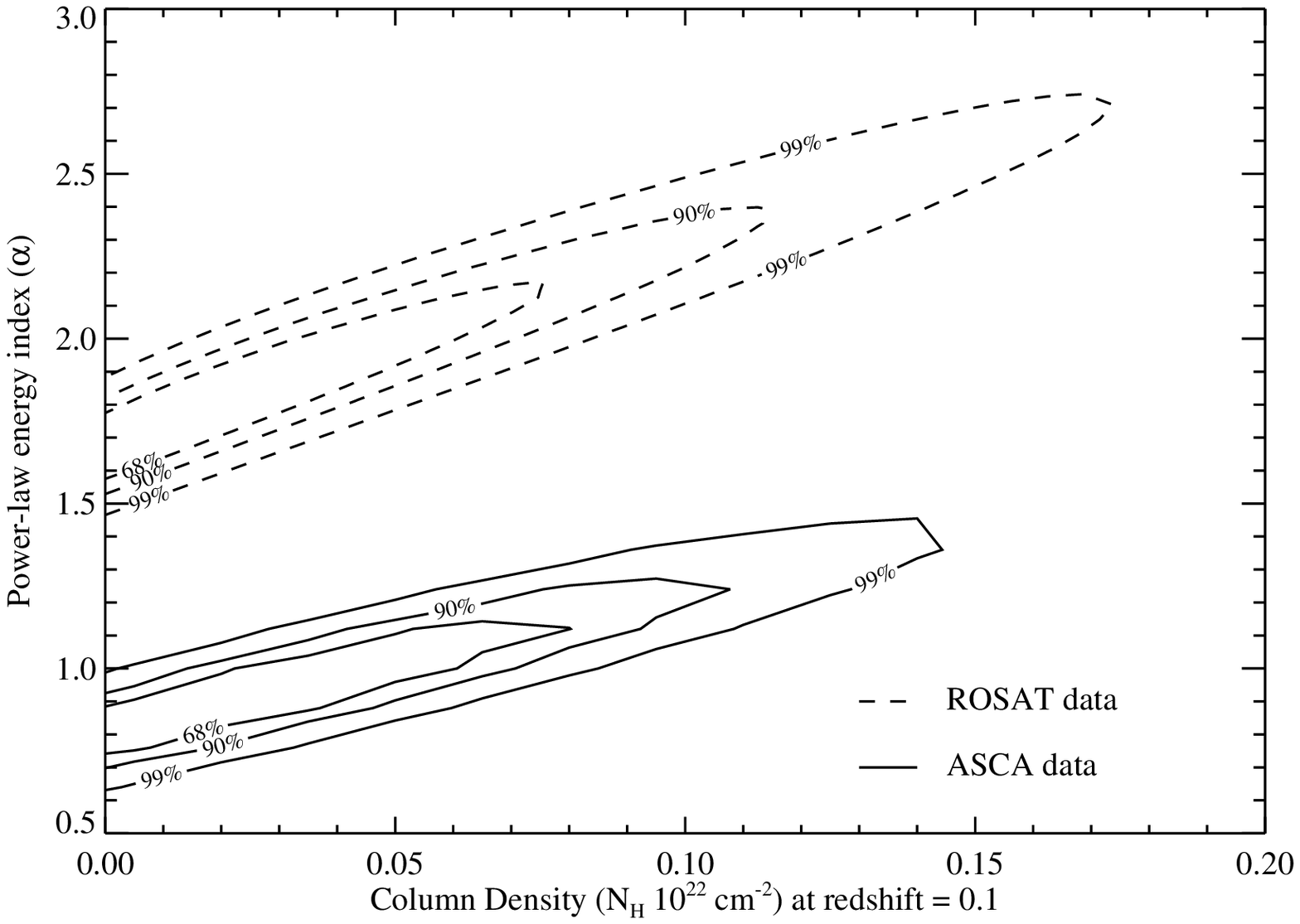,height=4.0in,width=6.5in,angle=0}
\caption{{\bf Figure 2.} Confidence contours at the 68, 90 and 99 per cent
level for the \ROSAT\ PSPC and \ASCA\ SIS data fitted independently to single
power-law models, over energy range 0.5--2\keV, showing that the two data sets
are different at more than 99 per cent significance. Note however that the
column density is consistent for both data sets while the power-law index is
different.}
\endfigure

\beginfigure*{3}
\psfig{figure=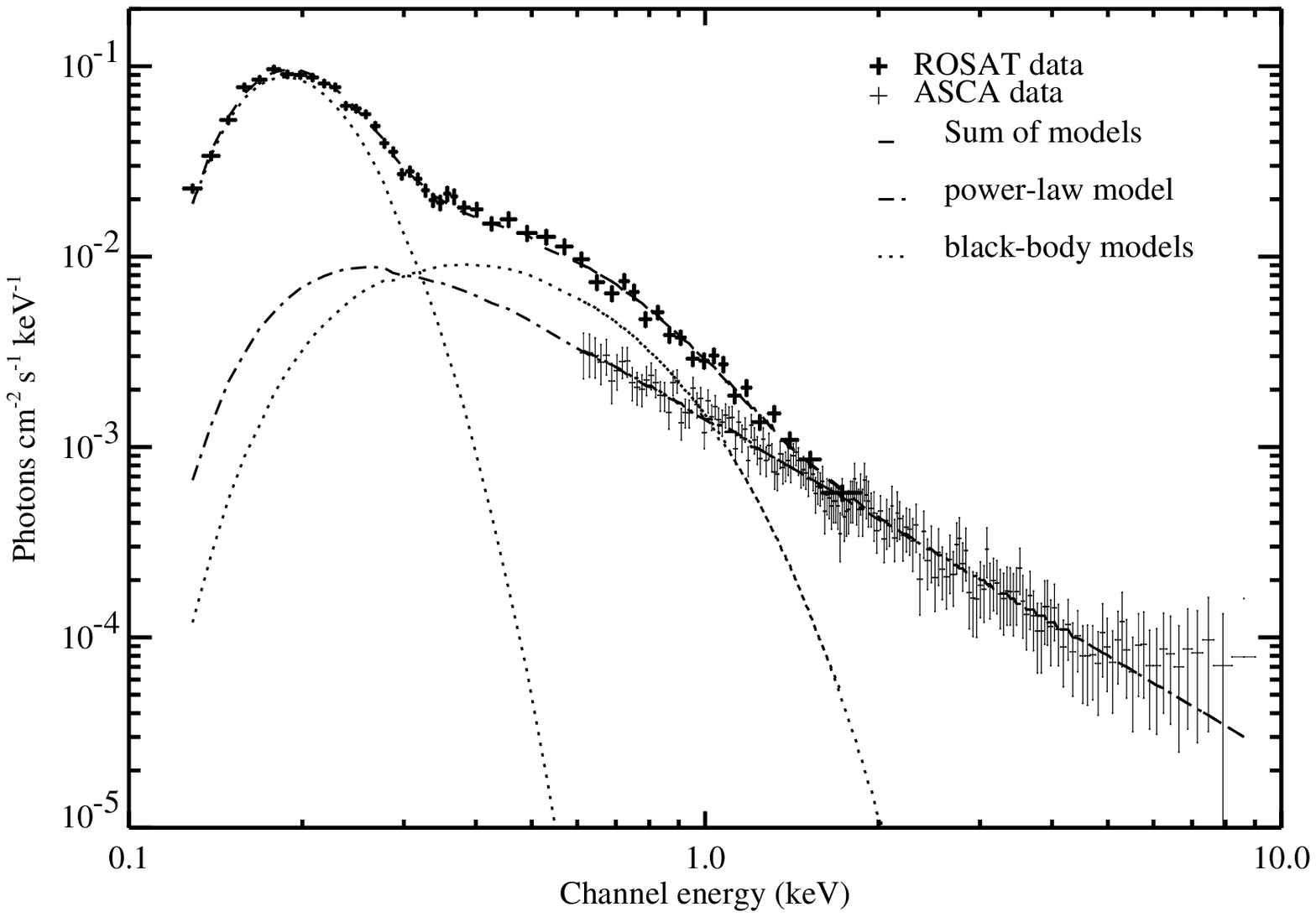,height=4.0in,width=6.5in,angle=0}
\caption{{\bf Figure 3.} The \ASCA\ data is fitted to a single power-law model
while the \ROSAT\ PSPC data is fitted to a composite model consisting of the
\ASCA\ power law, plus two black bodies.}
\endfigure

\beginfigure*{4}
\psfig{figure=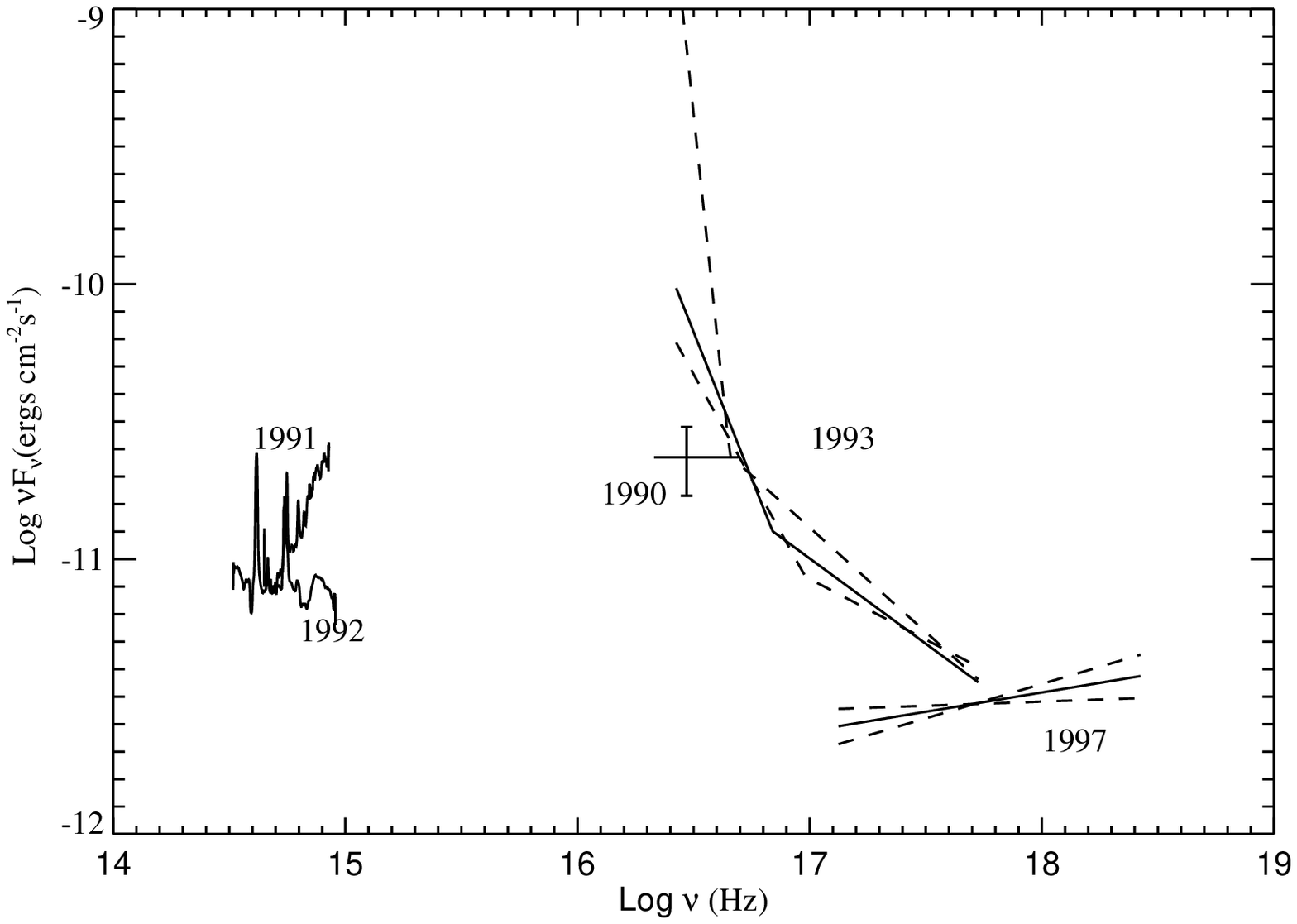,height=4.0in,width=6.5in,angle=0}
\caption{{\bf Figure 4.} This multiwavelength spectrum includes: the WFC RASS
observation and the pointed PSPC data (P95) plotted as a broken power law; the
\ASCA\ data  described here plotted as a single power law; the optical data
(M95); and the optical data taken by G98.}
\endfigure

\subsubsection{Short term variability}

During the pointed \ROSAT\ observation, the flux of \break \2248\ fell by 14
per cent in 16 hours (P95). The mean \ASCA\ count rates for SIS0, SIS1, GIS2
and GIS3 respectively were $0.209\pm 0.006$, $0.156\pm 0.004$, $0.106\pm
0.003$, $0.123\pm 0.003$\cs. Light curves were prepared from these \ASCA\ data
with bin sizes of 204, 264, 388 and 335~s respectively, to give a minimum of 50
counts per bin. The deviation from constant was given by $\chi^2$/dof = 
164/107, 98/79, 34/53 and 73/63 for SIS0, SIS1, GIS2 and GIS3 respectively.
Thus, there is no significant variability observed in the light curves of any
of the detectors, over a period of $\sim 17$ hours. 

\subsection{Long term variability}

\subsubsection{Comparison between \ROSAT\ PSPC, WFC and \ASCA\ data}

The pointed \ROSAT\ PSPC data taken in 1993 were shown by P95 to have a higher
soft X-ray excess than that expected from the original WFC observation,
suggesting a variable soft X-ray component. To test for further evidence of
long term variability in the soft X-ray component, we re-reduced the \ROSAT\ 
PSPC data using the same method as P95, and we find results that agree with the
best fitting model given in that paper (see Section 1). P95 examined the
possibility that there could be a `hole' in the galactic hydrogen column
(\nHgal), such that the measured value was greater than the true value, making
the ultrasoft component appear stronger than it really was. However, P95 were
able to reject this hypothesis by comparison of several \nHgal\ measurements
and also IRAS data.

The power law which had given a good fit to the \ASCA\ SIS combined data failed
to reproduce the PSPC data at greater than 99 per cent confidence. Simultaneous
fits to the SIS combined data and the PSPC were unsatisfactory:  both a broken
power law and a black body combined with a power law converged to  \Xv2$=2.8$. 

To assess the extent to which the spectrum had changed, the PSPC and \ASCA\
data were compared only in the overlapping energy range, 0.5--2.0\keV.  The
PSPC and SIS spectra were each fitted with an absorbed power law, with the
local \nHint\ allowed to go free, and the two sets of confidence contours are
plotted in Fig. 2. This shows that the 0.5--2.0\keV\ spectra are different at
more than 99 per cent significance, although the Hydrogen column is consistent
with being the same for both data sets. These data were then fitted
simultaneously, constraining \nHint\ to be the same for both data sets  while
freezing \nHgal\ at $1.4 \exp{20}$\cm2, but allowing different power-law slopes
and normalizations. The resulting  intrinsic Hydrogen column was $1.5 \exp{20}
$\cm2 and the PSPC spectrum was soft with $\alpha = -1.8$, whereas the SIS
data were hard with $\alpha = -0.9$. Table 2 gives the results from the fitting
of SIS and PSPC data separately and simultaneously. Thus it appears that the
0.5--2\keV\ spectrum was significantly harder when the \ASCA\ data were taken. 

In fitting the 0.5--2\keV\ PSPC data, it should be noted that even this
restricted band has some contamination from lower energy emission due to the
limited energy resolution of the PSPC ($\Delta E / E = 0.4/\sqrt{(E/keV)}$;
Walter \& Fink 1993). In order to test whether the contamination could disguise
consistency between the PSPC and \ASCA\ data, we simulated the PSPC data as a
broken power law, assuming a slope consistent with \ASCA\ ($-0.9$) above
0.5\keV. The amount of `contamination' was gradually increased by adjusting the
slope of the lower energy power law until a single power law (0.5--2.0\keV)
fitted to this simulation gave a slope of $-1.4$, which would be just
consistent with the real PSPC fit at the 1 percent confidence level (see Figure
2). The simulated lower energy power law required a slope steeper than $-6.0$ to
reproduce the real fitted parameters, and this is much steeper 
than is observed.

Spectral indices as measured by the PSPC are believed to be systematically
steeper than those measured by other instruments in overlapping wavelength
ranges. The difference has been estimated to be between 0.2 (Fiore \etal 1994)
and 0.4 (Iwasawa, Brandt \& Fabian 1998, Iwasawa, Fabian \& Nandra 1999),
depending on the shape of the spectrum being measured. The possible systematic
error of $2 \exp{20}$\cm2 on the SIS measurement of \nHint\ described in
Section 2.2 may also affect the comparison. As an apparent column density less
than this value is probably not significant, the SIS fit in Table 2 was
repeated with \nHint\ forced to be zero. This resulted in a slightly harder
power-law slope of $-0.88$ and no significant change to the other values.  The
power-law fit in Table 2 was also repeated with \nHint\ $2 \exp{20}$\cm2 less
for the SIS data than the PSPC data. This led to the two slopes getting
slightly closer together with  $\alpha$ changing from $-1.8$ to $-1.7$ for the
PSPC and from $-0.9$ to $-0.95$ for the SIS. By examining the contour plot in
Fig 2, it can be seen that with this change in the slope of the SIS data, \it
and \rm a reduction in the PSPC slope of 0.5, there is only just an  overlap at
the 99 percent level. Thus, even including the most pessimistic systematic
errors on both the PSPC and SIS data, it is very unlikely that the PSPC and SIS
measurement can be consistent with one another.

It may be that the change in the spectral slope between the PSPC and \ASCA\
spectra is due to variability in a soft component only, while the hard
component stayed constant throughout.  To test this, we re-fitted the PSPC data
assuming that a power-law component, consistent with that measured from the
\ASCA\ data, extended through the ROSAT range.  Various soft components were
added to complete the model fit to the PSPC  data.  The power law (slope, flux
and intrinsic column density) for the hard component was fixed from fitting to
the SIS data (for the 0.5--10\keV\ range; \nHint$= 0$\cm2 and
$\alpha = -0.83$; see Table 1) and the additional soft components tried
included blackbodies, Bremsstrahlung components and additional power laws. The
best fitting model (\Xv2 = 1.2) was obtained by adding a combination of two
black bodies, one at $kT_{\rm bb}$=28\eV\ and the other at $kT_{\rm
bb}$=160\eV\ (see Table 2). 

Fig. 3 shows both the SIS and the PSPC spectra together with the underlying
power law, which  fits the SIS data on its own, and the two additional black
body components which are required to fit the PSPC data. This illustrates the
nature of the variability of the soft X-ray spectrum. The \ASCA\ data cannot be
used to constrain the presence of the lower temperature (28\eV) black body, but
an upper limit on the flux of the higher temperature (160\eV) black body is 10
times less than the ROSAT fit, at 99 per cent certainty. The flux in the higher
temperature black body is $4.3\exp{-12} $ ergs\cm2\persec (luminosity 
$2\exp{44} $ ergs\persec), thus a change in flux of at least  $3.9\exp{-12} $
ergs\cm2\persec\ has been observed. If the whole of the lower temperature black
body has also disappeared, this gives an upper limit on the flux change of
$7.2\exp{-12} $ ergs\cm2\persec. For comparison, the flux in the power law
(2--10\keV) is $4.6\exp{-12} $  ergs\cm2\persec (luminosity $2.2\exp{44} $
ergs\persec).

The original WFC RASS observation with a count rate of $18\pm 5$\cks\ in
detector S1a (covering the band 90--206\eV; Wells \etal 1990) was found by P95
to be inconsistent with the later PSPC observation, from which the best fitting
power-law model predicted a WFC count rate of $38\pm 6$\cks. The RASS PSPC data
have been described by G98 as having a slope of $\alpha = -1.9\pm 0.1$ with a
negative \nH\ (\nHint$\sim -0.5\exp{20}$\cm2). If this were modelled with the
Galactic \nH, it would require a softening at the lowest energies. Using the
model of G98 we predict a count rate of $29\pm 4$\cks\ in S1a of the WFC, which
is higher than that measured, but not by more than 2$\sigma\ $ significance, 
which is allowable given the calibration uncertainties at the lowest energies
in the PSPC. No direct comparison can be made between the WFC data and the
\ASCA\ data because there is no overlap between the detectors. Extrapolating
the \ASCA\ spectrum to the WFC range however gives a prediction of less than
1\cks. 

\begintable*{2} 
   \caption{{\bf Table 2.} Spectral fits to ASCA SIS and ROSAT PSPC data}
   \halign{#\hfil\quad & #\hfil\quad &
   	\hfil#\hfil\quad & 
   	\hfil#\hfil\quad & 
   	\hfil#\hfil\quad & 
   	\hfil#\hfil\quad & 
   	\hfil#\hfil\quad & 
   	\hfil#\hfil\quad & 
   	\hfil#\hfil\cr
   Model   	& Range (keV)	& \nHint$^a$ 		& $\alpha$ 	
		& Norm$^b$	& kT$_{\rm bb}$ (eV)	& Norm$_{\rm bb}^b$  	
		& \X2 /dof 	& \Xv2 \cr 
\cr
   PSPC  	& 0.5--2.0 	& $1.15^{+10.4}_{-4.0}$ & $-1.8^{+0.4}_{-0.6}$
		& $2.9\pm0.3$ 	&			&
		& $25.2/19$	& 1.3 \cr 
\cr
   SIS  	& 0.5--2.0 	& $2.2^{+8.3}_{-5.4}$ 	& $-0.98\pm 0.34$ 
		& $1.6\pm0.07$ 	&			& 	
		& $58/96$	& 0.60 \cr 
\cr
   power law$^c$& 0.5--2.0 	& $1.5^{+6.4}_{-4}$ 	& $-1.8 \pm 0.15$ (P) 	
		& $2.9\pm0.1$ (P) &			& 
		& $83.5/116$	& 0.71 \cr 
		&  		&  			& $-0.9\pm 0.13$ (S) 	
		& $1.5\pm0.5$ (S)  &			&
		&		& \cr 
\cr
   PSPC		& 0.1-2.0 	& $0$			& $-0.83$
		& $1.5$		& $28\pm 1.2$		& $0.76\pm 1.8$
		& $57.85/48$ 	& 1.2  \cr
   PL + 2bb$^d$	&		&			&
		&		& $160 \pm 6$		& $0.06\pm 0.08$	
		&		&	\cr
\cr
    } 
   \tabletext{A value of $1.4 \exp{20}$\cm2 has been assumed for the
Galactic column density.

\noindent (P) and (S) indicate values associated with the PSPC or SIS only,
respectively.

\noindent `bb' is used to indicate black body.

\noindent $^a \exp{20}$\cm2

\noindent $^b \exp{-3}$ photons \cm2\persec\ at 1\keV

\noindent $^c$ fitting the SIS and PSPC data simultaneously, but allowing the
spectral slope and normalization to vary independently.

\noindent $^d$ PSPC data fitted to a fixed power law and normalization
determined from the SIS (see Table 1) plus an additional two black bodies. 

}
\endtable

\subsubsection{Variation in optical data}

G98 describe optical spectroscopy of \2248\ taken in 1992. These data were
taken in photometric conditions with a 2 arcsec slit.  Whilst they were not
taken at the parallactic angle, the zenith distance of the object was high when
the spectrum was taken and a comparison with CCD images shows that no blue
light has been lost through refraction (H.-C. Thomas, private communication,
1998). G98 measure a slope in the optical of $\alpha = -1.3$, much softer (\ie
redder) than the $\alpha = +0.8$ measured from the 1991 M95 data. We have
plotted their optical data in Fig. 4 for direct comparison with the M95 data.
Assuming a conservative systematic error of 30 per cent on the M95 spectrum, we
find that the fluxes in the two optical spectra are not inconsistent, although
the continuum slope has changed dramatically. In the two spectra,  the
continuum fluxes are approximately equal at the wavelength of \Hb, and the
emission line FWHM, EW and fluxes of the \Hb\ lines are very similar. Whilst
the FWHM and  fluxes of the \Ha\ lines are also similar, the \Ha\ EW is higher
in the M95 data, and is associated with a lower continuum level at this
wavelength. However, given the possible errors on the absolute flux
measurements, we cannot say for certain whether the line fluxes have stayed
constant.

\subsubsection{Multiwavelength spectrum}

Fig. 4 shows the multiwavelength spectrum including the pointed PSPC and
\ASCA\ data together with the optical data. The WFC point is also included.  
Significant changes in the overall optical to X-ray spectrum are obvious. In
X-rays, one interpretation could be that the soft component was strong in 1990,
steepened further at the lowest energies in 1993 but then fell away completely
in 1997 (note, though, the caveats in Section 2.3.1). Meanwhile, the BBB was
unusually strong during 1991, yet by 1992 there was little or no emission from
this component. These data suggest many intriguing possibilities for the
behaviour of the optical to X-ray spectrum in \2248. For example, if the BBB
and the ultrasoft component are part of one `big bump', the flux in this
component may be changing dramatically from very low levels to where it
dominates the spectrum completely. Alternatively, the big bump may be shifting
in frequency, \ie changing its characteristic temperature significantly. It is
also possible that the BBB and ultrasoft component are independent features
and there is no commonality in their behaviour. 


%
%
%

\section{Discussion}

Optical and X-ray spectra of \2248\ have now been measured over a period of
seven years and reveal extreme changes in both ranges. In this paper, we
present the results from a new \ASCA\ measurement and compare it with previous
\ROSAT\ (PSPC and WFC) and optical data. 

The pointed \2248\ PSPC data, modelled with a single power-law in the full
0.1--2\keV\ range (for fixed Galactic $N_{\rm H}$) gives a very poor fit (P95).
It is necessary to use a broken power law with a soft index (below 0.26\keV) of
--3.13 and a hard index --1.62. The \ASCA\ X-ray spectrum (0.5--10\keV), on the
other hand, is well fitted by an $\alpha = -0.85\pm 0.08$ power law with no
evidence for any curvature, soft excess, or excess absorption. This is
consistent with the mean \ASCA\ energy index determined for broad line Seyfert
1s by Brandt \etal (1997) who measured $-0.87\pm 0.036$. Similarly, Comastri
\etal (1992) found a mean 2--10\keV\ slope of $-0.89\pm 0.06$, with most
objects distributed around $\alpha \sim -1.0$, from an \EXOSAT\ sample of AGN.
Thus despite the unusual two-component low energy X-ray spectrum measured by
the PSPC in 1993, \2248\ had a high energy X-ray continuum typical of AGN in
general at the time of the \ASCA\ observation, with no sign of any soft X-ray
excess.

In Section 2.3.1, we showed that the changes between the PSPC and the \ASCA\ 
spectrum observed were consistent with a changing soft X-ray spectrum (modelled
by two blackbodies, $kT_{\rm eff}=28$\eV\ and 160\eV) superposed on an
invariant, underlying $\alpha=-0.8$ power law. However, while the lack of any
`soft X-ray' component emission in the \ASCA\ spectrum is ably demonstrated by
these data, the slope of any underlying power-law in the earlier PSPC spectrum
is very poorly constrained. The `ultrasoft' component in the PSPC spectrum
(modelled by the 28\eV\ black body) is below the spectral range covered by
\ASCA\ and therefore cannot be compared with the later spectrum. Note, however,
that we have no evidence that the `ultrasoft' and `soft' X-ray components are
in fact separate components; produced by different mechanisms and varying
independently. We can only say that the shape of the spectrum does not conform
to a single power law or single blackbody.

\subsection{\2248\ compared with PG quasars}

With an absolute magnitude $M_{\rm B} < -23$, \2248\ meets the definition of a
quasar given by Schmidt \& Green (1983) for the PG Bright Quasar survey. At $z
= 0.101$, and with a Galactic column density \nHgal$ =1.4\exp{20} $\cm2, it is 
directly comparable with the L97 sample of PG  quasars\note {$^\star$}{L97
selects bright PG quasars with low redshift ($z\le 0.4$) and local Galactic
column density  (\nHgal\ $< 1.9 \exp{20}$\cm2)}. L97 found no evidence for
spectral curvature in their sample and  the L97 mean spectral index in the
\ROSAT\ PSPC energy range (0.1--2\keV) is  $\alpha\sim -1.7 \pm 0.1$.  

If we consider only the slope of the  \2248\ `soft X-ray' component above
0.3\keV, in both the pointed and the RASS \2248\ PSPC observations, it appears
typical of the L97 sample.  Similarly, given the slope of the 0.3--2\keV\
spectrum, the optical to X-ray ratio of \2248\ is consistent with their sample
with  $\alpha_{\rm ox} =  -1.3$ (using the definition of \aox\ of L97 \ie the
slope between 3000\AA\ and 2\keV). In other words, \2248\ exhibits an otherwise
flat, normal, underlying optical to X-ray continuum; what sets this object
apart is the apparent superposition of an unusually strong ultrasoft
($<$0.3\keV) component. 

Most of the optical slopes from L97 (\ao) are small and negative
(with a mean of $-0.36\pm 0.22$). None are as hard (blue) as the 1991
measurement of  \2248, when an $\alpha_{\rm opt} \sim 0.8$ was observed. In
1992 however, the index of the optical continuum fell to --1.3, much {\sl
softer} (redder) than the L97 quasars; another dramatic change in the
shape of the continuum of \2248.

\subsection{The X-ray spectrum and the BLR}

Before any investigation of the relationship between the BLR and the soft X-ray component can
be made, it is essential to establish the phenomenology of the AGN, ie. whether or not it can
be described as `ultrasoft'. In Boller \etal (1996), the average slope, quantified by a single
power-law fit to the 0.1 to 2.4\keV\ {\sl ROSAT} PSPC band, was used to assess the softness of
the Seyfert galaxies. However, it  does not take into account any significant curvature in the
spectrum, so that AGN like \2248, whose PSPC spectrum rises steeply below 0.3\keV\ (P95), may
have been overlooked.  For example, when the PSPC spectrum of \2248 is fitted with a single
power law (where the local column density is fixed to the Galactic column), the power-law index
$\alpha$ converges to -1.8, which is not particularly soft when compared with other AGN. Taking
this index together with the Balmer line FWHM of $\sim$3000~km~s$^{-1}$, the object does not
lie in the zone of avoidance as described by Fig. 8 in Boller \etal (1996). However, with a
$\chi_\nu^2$ of 2.4 (for 49 degrees of freedom), a single power law is an unacceptable fit to
the spectrum. A concave broken power law provided the best fit, with a low energy slope
($<$0.3\keV) of $\alpha$=--3.2 (a $\chi_\nu^2$=1 for 49 degrees of freedom), thus model fitting
indicates that \2248 does have a very steep, soft  X-ray excess.

Furthermore, the original definition of an ultrasoft source was made by
Puchnarewicz \etal (1992) and C\'ordova \etal (1992). An ultrasoft source had
to satisfy the criterion that the C1 to C2 count rate ratio 
must exceed 2.8, where C1
covers the 0.16 to 0.56\keV\ band of the {\sl Einstein} IPC detector and C2
covers the 0.56 to 1.08\keV\ range. By folding the best-fitting PSPC model
through the IPC response, we find that the C1 to C2 ratio for \2248 is 3.1;
thus satisfying the criterion for an ultrasoft AGN.

We find, therefore, that \2248 is a variable, ultrasoft X-ray AGN with broad
lines (Balmer line FWHM $\sim$3000~km~s$^{-1}$). The presence of broad lines
is unusual for an ultrasoft AGN (Puchnarewicz \etal 1992),
consequently this source is very important in our understanding of the
relationship between the BLR velocity and the soft X-ray excess.

In 2.3.2 the two optical spectra of \2248\ were described. If the Balmer line
fluxes and FWHM have remained the same between 1991 and 1992, the data suggest
that the changes in the central ionizing continuum have not affected either 
the velocity of the outer BLR, nor the line strengths, at least not on a
timescale of about a year.

Baldwin \etal (1995) showed that efficient line emission occurs where the
combination of conditions in the gas and the nature of the ionizing continuum
are optimized. Thus it is possible that clouds may cover a wide region but are
only `lit-up' where these conditions are right. If the FWHM of the Balmer lines
in \2248\ {\sl have} remained unchanged throughout the very large variations in
ionizing continuum strength and shape, this would suggest either that this
fine-tuning takes a long time to respond (more than a year), or that there is
no low-velocity gas in \2248.

\subsubsection{The big bump and the soft X-ray spectrum}

It has been suggested that the soft X-ray excess is the upper energy tail of
the big blue bump, although studies have been unable to prove a direct
relationship between these two components. However, the correlation between
\ax\ and the Balmer line FWHM links the velocity of the outer broad line region
with the shape of the soft X-ray spectrum. \2248\ is consistent with this trend
{\sl if} the 0.3-2\keV\ part of the PSPC spectrum {\sl only} is considered. 

One major difference between \2248\ and other Seyfert 1s is the presence of the
ultrasoft component which is very variable. Another is the very hard (blue)
optical continuum, which is also very variable. We speculate that these unusual
X-ray and optical continua may be related, \ie that they may be part of a very
strong and variable big bump. The Balmer line region velocity is not affected
by the presence of the ultrasoft component but {\sl is} consistent with the
slope of the soft X-ray component. However, since the \aox\ for \2248\ is
typical of the L97 quasars given its \ax, some relationship between the big
bump and the soft X-ray spectrum cannot be ruled out.

\subsection{Comparison with other ultrasoft AGN}

There are several other ultrasoft AGN which have shown dramatic variability in
the soft component, such as \break RE~J1237+264, (Brandt, Pounds \& Fink 1995)
and WPVS 007 (Grupe \etal 1995). These objects are both NLS1s and showed a
reduction in the soft component by factors of 70 and 400 respectively.
E1615+061 (Piro \etal 1988), however, is an ultrasoft, broad line
object which showed variability such that it became less soft when it became
less luminous. 

The object \H\ (Marshall, Fruscione \& Carone 1995, Fruscione 1996) shows
remarkably similar characteristics to \2248\ (Guainazzi \etal 1998). It is
categorized as a Seyfert 1.5 with a redshift of 0.103 and has a strong, broad
\Hb\ component of FWHM 3500\kms.  The PSPC data are not well fitted by a simple
power law due to a softening at low energies, and require a broken power law,
or a power law plus an additional soft component (\ie black body,
Bremsstrahlung or another power law). The blackbody temperature of the
ultrasoft component is around 40\eV. A comparison of \ROSAT\ PSPC, \ASCA\ and
\SAX\ data revealed a highly variable soft component and little evidence of
iron lines at around 6.4--6.7\keV.  

\0437\ (Wang \etal 1997) is another ultrasoft Seyfert 1 (z = 0.051) whose PSPC
data cannot be described by a single power law, but requires an ultrasoft
component. This component can be described by a black body with a temperature
of 22\eV. Large amplitude flux variability has been observed in the \ASCA\
range. This object differs from \2248\ in that it has an unusually steep \ASCA\
slope for a broad line object (\Hb $\sim 4300$ \kms) with $\alpha \sim -1.2$. 

\subsection{Models of the X-ray spectrum}

\subsubsection{Accretion discs}

Comptonized accretion disc models have been used successfully to model the
continuum production in AGN. For example, the hot corona model proposed by
Haardt \&  Maraschi (1993), describes a feedback system whereby medium energy
X-rays from a hot, optically-thin corona are reprocessed by a cold, dense
accretion disk into soft blackbody photons. Some of these blackbody photons 
become seeds for Comptonization in the hot corona, forming a medium to high
energy X-ray continuum (see also Haardt, Maraschi \& Ghisellini 1997). Their
models can simulate systems in which the X-ray power-law slope changes in
response to changes in flux from the source, and changes in opacity in the
corona. The spectral changes observed in both \H\ and \2248\ can be explained
by their  model, by invoking a change in opacity ($\tau$) from $\sim 1$ to
$\sim 0.1$. The corona cannot be electron-positron pair dominated, because this
would require even a small change in spectral index to be accompanied by a
change in flux larger than a factor of 2, which is not observed. Guainazzi
\etal (1998) discuss the Haardt \etal (1997) model with respect to \H, and also
viscousless, two-phase shock accretion disk models by Chakrabarti \& Titarchuk
(1995) and Ebisawa, Titarchuk \& Chakrabarti (1996). In the latter model, the
changes in the \H\ spectrum are assigned to a transition from a bulk motion to
a thermal motion regime due to a change in accretion rate.

Pounds, Done \& Osborne (1995) have suggested that the NLS1 \1034\ is a
high-$M$ analogue of a Galactic Black Hole Candidate (GBHC) in a high state. In
this state,  the ultrasoft component is steep and soft, while the hard
component is very weak and variable but is also soft relative to the low-state
slope in hard X-rays. Further evidence for this GBHC model was given by Brandt
\etal (1997), who found that the \ASCA\ slopes of NLS1s in general are soft
relative to broad line Seyfert 1s (see also G98). With the ultrasoft component
in \2248\ switching from a high to a low state, it could also be a candidate
for the GBHC analogy, although one that fits uneasily into the NLS1 scenario
because of its broad lines. However, the disappearance of the soft component
and hardening of the X-ray power-law are consistent with the behaviour of
GBHCs. Further investigation of this model would require simultaneous
monitoring of the soft and hard X-ray spectra (\ie from 0.1 to $\sim$10\keV\ or
higher) however; the present data are insufficient.

\subsubsection{Reprocessing models}

Collin-Souffrin \etal (1996) found that the near-IR to X-ray spectra of AGN can
be modelled by a system of radiatively heated clouds, optically thick to
electron scattering, with a covering factor greater than 0.5.   These clouds
could be in a `quasi-spherical' distribution or a thick disk. The X-ray
spectrum produced by this model is a combination of the primary incident
continuum (assumed to be a featureless $\alpha=1$ power law) and the component
reflected from the hot, illuminated side of the clouds. (The optical/UV
emission is produced by the cooler, unilluminated side of the clouds.) In order
to predict an ultrasoft excess of the strength observed in \2248, the spectrum
must be dominated by the reflection component. However, in this case the
FeK$\alpha$ line would also be strong, yet only an upper limit could be
measured from the \ASCA\ data of \2248. Variability studies would provide
further evidence relevant for this model since it predicts that the UV should
respond on longer timescales than X-rays, although which wavelength leads will
depend on the geometry of the system.

\section{Conclusions} 

\2248\ has displayed a most unusual low energy X-ray spectrum for any type 1
AGN. A steep, variable, ultrasoft component is found below 0.3\keV, which
appeared from ealier \ROSAT\ data to be superposed on a relatively normal soft
X-ray spectrum. \2248\ has also exhibited an extremely blue, but variable,
optical spectrum. The broad Balmer line widths are consistent with \ax\ above
0.3\keV\ and appear to be unaffected by the strength of the big blue bump. 

The \ASCA\ spectrum presented in this paper has shown the soft X-ray component
(0.5--2\keV) also to be highly variable, falling away to the extent that it is
no longer detectable. The hard X-ray (0.5-10\keV) continuum can be fitted with
a single, absorbed power law of slope $\alpha\sim -0.85$ which shows no
indication of short timescale variability. Neither does it show significant
evidence for absorption edges or emission lines. It may be extrapolated through
the \ROSAT\ PSPC spectrum taken 4 years earlier, if the soft excess measured
then is modelled by two blackbodies ($kT_{\rm eff} = 28$\eV\ and 160\eV).

The spectral changes are similar to those observed in GBHCs as those systems
fall into a low from a high state. In the high state, the ultrasoft component
is strong and the hard X-ray continuum is soft, while in the low state the
ultrasoft component falls away and the hard power-law hardens. The former
spectrum was measured for \2248\ in 1993 by the PSPC (although the slope of the
hard X-ray power law was poorly constrained due to the limited range of the
PSPC) and the latter by \ASCA\ in 1997. Even taking account of pessimistic
systematic and cross-calibration errors, the changes between the two spectra
are significant. An alternative model to the GBHC analogy is that the opacity
in a hot, optically-thin corona surrounding an accretion disc has fallen. 

The extreme patterns of variability observed in the X-ray and optical spectra
of \2248\ reveal a fascinating, highly dynamic source which can reveal much
about the physics in the central regions and their relationship with the outer 
line-emitting gas. With the sparse, non-simultaneous data in hand, firm
conclusions are difficult to draw. Optical and X-ray monitoring are essential
to proceed further with investigations of this AGN.

\section*{Acknowledgments}

We thank Keith Mason for his advice. We also thank D. Grupe and H-C. Thomas for
allowing us to print their optical spectrum, and for useful comments on the
paper. We would also like to thank the referee for some constructive and useful
comments leading to an improved paper.

\bye